\documentclass[a4paper]{article}

\usepackage[english]{babel}
\usepackage{shuffle}
\usepackage{amsfonts}
\usepackage{amsmath, amssymb}
\usepackage{mathtools,calc}
\usepackage[]{cite}
\usepackage[margin=1in]{geometry}
\usepackage{hyperref}
\usepackage{physics}
\newcommand{\e}[1]{\mathrm{e}^{#1}}
\DeclareMathOperator{\sop}{\Sigma}
\DeclareMathOperator{\lop}{\Lambda}
\numberwithin{equation}{section}

\newcommand{\orcid}[1]{\href{https://orcid.org/#1}{\includegraphics[width=8pt]{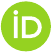}}}

\title{A permutation-based power series representation of the Baker-Campbell-Hausdorff formula}

\author{Joseph M. Jones\thanks{jxj898@theory.bham.ac.uk}
	\orcid{0009-0005-0709-8615} and M. W. Long}
	
\begin{document}
\maketitle
\begin{abstract}
	The Baker-Campbell-Hausdorff formula was recently resummed exactly in one variable, and left as a power series in the other (Moodie and Long 2021 \textit{J. Phys. A: Math. Theor.} 54 015208).
	The coefficients of the power series were provided as a sum of products of three hyperbolic functions that are analogous to the familiar commutator expansion. We find a new form of the power series coefficients that is a linear combination of just one of the hyperbolic functions. This linear combination can be understood through elementary permutations of the arguments of the hyperbolic function. We use generating functions and hyperbolic identities to relate the representations.
	The permutation representation is radically different to previously known structures in the Baker-Campbell-Hausdorff formula and naturally supersedes the previous power series representation for future use, in our opinion.
	It also allows us to prove a set of intriguing \textit{marching identities} that are useful in physical applications and were discovered in the initial work.
\end{abstract}
\section{Introduction}
A power series representation of the Baker-Campbell-Hausdorff (BCH) formula was recently found by one of the authors and presented in a paper \cite{moodie2021} that we will refer to as \textit{the previous paper}. The previous authors argue that their power series provides a better approximation of the full BCH formula upon truncation since one of the variables has been resummed exactly. The remaining series is perturbative in only one variable unlike the usual BCH commutator formula \cite{campbell1897,baker1905,hausdorff1906,dynkin1949} where both parameters are assumed to be small \cite{achilles2012} for one to expect truncation to be a good approximation. The case when one variable is small is particularly relevant to classical statistical mechanics and one can derive a new perturbation theory using the power series representation to investigate phase transitions, the correlation length and its associated critical exponent. This was the physical motivation for this work and will be the subject of future work \cite{jones2025}. A preliminary example of how this formula may be used is provided in Appendix~\ref{app:pert}.
The coefficients of the power series are the quantities of interest in this paper and we refer to the previous paper's result as the \textit{original representation}. They satisfy what we refer to as \textit{marching identities} which are proven later and are crucial to deriving perturbation theory; this was the mathematical motivation for the present paper.

The previous authors considered a symmetric form of the BCH identity,
\begin{align*}
	\e{2C}=\e{A}\e{2B}\e{A}
\end{align*}
and proved that it has the power series representation
\begin{align}
	C &= A +\sum_{N=1}^\infty G_N(L_1,L_2,\dots,L_N) B_1B_2\cdots B_N\label{eq:ml-series},
\end{align}
where the $G_N$ are hyperbolic functions that are known to all orders, $L B\equiv[A, B]$, and we have removed an irrelevant factor of two from the previous result. The $G_N$ were defined in terms of the original representation, which is a sum of products of special functions $f_r$ (see equations~\eqref{eq:fh},~\eqref{eq:h0} to~\eqref{eq:h6} and surrounding discussion for details). Explicit details of $G_N$ are provided later when they are relevant. The subscript on $L_m$ specifies which $B$ in the ordered product it acts on, for example 
\begin{align*}
	L_mB_1B_2\cdots B_{m-1}B_mB_{m+1}\cdots B_{N-1}B_N=B_1B_2\cdots B_{m-1}[A,B_m]B_{m+1}\cdots B_{N-1}B_N.
\end{align*}
The subject of the current paper is alternative representations of $G_N$ that each have different properties.
In physics one is often interested in the eigenvalues of an operator and those provided by equation~\eqref{eq:ml-series} are the same as for the more common formulation since they are related through a similarity transform, $\e{A}\e{2C}\e{-A}=\e{2A}\e{2B}\equiv\e{2D}$.

At first glance, the $G_N$ appear fully determined as they are provided explicitly, leaving no obvious freedom to create a new representation. 
However, the previous paper noted that $G_N$ separates into two halves related by
\begin{align}
	\sinh(L_1+\cdots+L_{N+1})\,G_{N+1}(L_1,\dots,L_{N+1})
	&= Q_N(L_1,L_2,\dots,L_N) - (-1)^{N+1} Q_N(L_{N+1},L_N,\dots,L_2) \nonumber\\
	&= (1 - (-1)^{N+1} R_{N+1}) Q_N(L_1,L_2,\dots,L_N), \label{eq:gn-rev}
\end{align}
where $R_{N+1}$ is the reversal permutation operator on $N+1$ arguments such that $R_NL_m=L_{N+1-m}$ and $R_{N+1}Q_{N}(L_1,L_2,\dots,L_{N})=Q_{N}(L_{N+1},L_{N},\dots,L_2)$ is the reversed copy of the first term.
Equation~\eqref{eq:gn-rev} shows that only a particular combination of terms is required to determine $G_{N+1}$, leaving freedom to modify $Q_N$ by any function $X_N$ satisfying
\begin{align}
	(1 - (-1)^{N+1} R_{N+1}) X_N(L_1,\dots,L_N) = 0. \label{eq:xn-rev}
\end{align}
Any such $X_N$ can be added without changing $G_{N+1}$.

The purpose of this paper is to fully exploit the reversal property introduced in the previous paper: we identify the extra function $X_N$ that relates our new representation to the original and verify that it satisfies equation~\eqref{eq:xn-rev}.

The first term in equation~\eqref{eq:gn-rev} depends on the first $N$ out of $N+1$ variables and the second term on the final $N$ variables. In this paper we relate various forms of $Q_N$ through generating functions then prove that equation~\eqref{eq:xn-rev} is satisfied for each $N$. Equation~\eqref{eq:xn-rev} implies that there is no $L_1$ dependence in $X_{N}$, and further that the dependence on $L_2$ is identical to that on $L_{N}$, that on $L_3$ is identical to that on $L_{N-1}$ and so on.

In the previous paper $Q_{N}$ was written as a sum over $N$ hyperbolic terms and a linear term, each hyperbolic term is proportional to a product of three hyperbolic functions, two that together include the arguments from $L_1$ up to $L_r$ and one that includes the rest from $L_{r+1}$ up to $L_{N}$. For example, a generic term in this representation is proportional to $f_r(L_r,\dots,L_2)f_1(L_1+\cdots+L_r)f_{N-r}(L_{r+1},\dots,L_{N})$, we refer to this as \textit{partition structure} where the partition is between variables $L_r$ and $L_{r+1}$ in this case. 
This original representation of $G_N$ is cumbersome and difficult to work with; it lacks both symmetric partitioning of the arguments and the underlying permutation structure. Our first step is therefore to use what we refer to as edge identities to combine the first two $f_r$'s to make a product of two $f_r$'s in total, giving a cleaner formulation of the original representation. We use this as the starting point to derive the permutation representation.

Throughout this paper, we progressively reduce the number of $f_r$ factors in our formulae. Section~\ref{sec:edge-K} lowers three to two, Section~\ref{sec:perm-rep} reduces four and three to two, and finally all terms can be expressed as sums over a single $f_r$ with permutation-structured arguments.

Now we state two equivalent forms of our main result that each have their own advantages, then explain the structure of the paper.
The first statement of the main result elucidates the formal mathematical structure, while the second is directly applicable to physical problems.

\subsection{Statement of main result}
We first require some basic permutation notation. The raising operator
\begin{align*}
	\sop\equiv(12)(23)(34)\cdots=\prod_{n=1}^{\infty}(n\,n+1)
\end{align*}
is core, controlling various quantities in this document, where $(n\, n+1)$ is the transposition of indices $n$ and $(n+1)$ and $\sop$ is defined on the infinite string of symbols, becoming a permutation operator if truncated to a finite product. For example $\sop$ behaves as 
\begin{align*}
	\sop L_m &=L_{m+1}\sop\\
	\sop g(L_1,L_2,\dots,L_m)&=g(L_2,L_3,L_4,L_5,\dots,L_{m+1})\sop	
\end{align*}
for some function $g$. 
The index $1$ is lost under the action of $\sop$ and so it is not simply a permutation.
Generally
\begin{align*}
	\sop \cdot = (\sop \cdot\sop^{-1})\sop
\end{align*}
so that the bracketed term has its arguments raised with the final $\sop$ acting on subsequent quantities and finally we define the crucial permutation operator
\begin{align*}
	\lop\equiv 1-\sop
\end{align*}
where $1$ represents the identity permutation, and $\lop$ generates the binomial structure by selecting either the identity or the raising operator at each application.
Note that $\sop$, $\lop$ and the identity permutation operator are defined on the $L_m$'s and not on the product of $B$'s, so that once they have acted and are to the right of all $L_m$'s they may be removed from expressions.

We can now state our main result; the symmetric form of the BCH identity has the power series representation
\begin{align}
	C&= A + \sum_{N=1}^{\infty}\frac{1}{\sinh(L_1+\cdots+L_N)}\lop H_{N-1}\,B_1\cdots B_N
\end{align}
where
\begin{align}
	H_N &= \sum_{p=0}^{N-1}t_p\lop^p\coth(L_1+\cdots+L_{N-p})\lop H_{N-1-p}+t_N\lop^{N}H_0,\quad	\lop H_0 \equiv L_1\label{eq:resH}
\end{align}
and $t_p\lop^p$ controls the gaps in the product of $\coth$'s (as shown in~\eqref{eq:H1-to-H4} and recognised in the previous paper), with $t_p$ being the coefficients in the Taylor series of $\tanh(x)/x$ as given in equation~\eqref{eq:tan-over-x}. The Taylor series of the hyperbolic functions produces the familiar commutator expansion of Dynkin \cite{dynkin1949,achilles2012}, as explained in the previous paper, and each term in the expansion~\eqref{eq:resH} represents an infinite number of terms in Dynkin's expansion.
The first few are
\begin{subequations}\label{eq:H1-to-H4}
	\begin{align}
		H_1 &= \coth(L_1)L_1\label{eq:H1}\\
		H_2 &=\coth(L_1+L_2)\lop\coth(L_1)L_1+t_2\lop L_1\\
		H_3 &= \coth(L_1+L_2+L_3)\lop\coth(L_1+L_2)\lop\coth(L_1)L_1\nonumber\\
		&\quad +t_2[\coth(L_1+L_2+L_3)\lop^2+\lop^2\coth(L_1)]L_1\\
		H_4 &= \coth(L_1+L_2+L_3+L_4)\lop\coth(L_1+L_2+L_3)\lop\coth(L_1+L_2)\lop\coth(L_1)L_1\nonumber\\
		&\quad +t_2[\coth(L_1+L_2+L_3+L_4)\lop\coth(L_1+L_2+L_3)\lop^2+\lop^2\coth(L_1+L_2)\lop\coth(L_1)\nonumber\\
		&\quad\quad+\coth(L_1+L_2+L_3+L_4)\lop^3\coth(L_1)]L_1+t_4\lop^3L_1\label{eq:H4}.
	\end{align}
\end{subequations}
Finally, the quantity 
\begin{subequations}\label{eq:omit-args}
	\begin{align}
		\lim_{\lop\to 1} H_N &=  h_N(L_1,L_1+L_2,\dots,L_1+L_2+\cdots+L_N)L_1\equiv  h_NL_1\\
		&=f_N(L_1,L_2,\dots,L_N)L_1\equiv f_NL_1\equiv[L_1L_2\cdots L_N]\label{eq:sym}
	\end{align}
\end{subequations}
is the central hyperbolic quantity in the later analysis, and the final line defines a shorthand symbol that is employed in the next statement of the main result and later in the proof of marching identities. 
The $h_r$ functions accumulate arguments relative to $f_r$ as specified in equation~\eqref{eq:fh}. The first few $h_r$ are given in equations~\eqref{eq:h0} to~\eqref{eq:h6}. Using $h_r$ in favour of $f_r$ is motivated by the fact that the analogue between $\coth(x)$ and $1/x$ is clearer in the $h_r$ representation, and this is extremely relevant to perturbation theory. When arguments are omitted from hyperbolic quantities of $n$ arguments they are always understood to be as in~\eqref{eq:omit-args}, apart from when we often require to specify the argument of a single $f_1(x)=h_1(x)=\coth(x)$.
Explicit low-order examples of the perturbation theory---which can be applied without appreciating the details of the proof---are supplied in Appendix~\ref{app:pert} for a reader concerned with physical applications.

\subsection{Permutation statement of main result}
An alternative form of the main result---that we use later in proving marching identities---is found by commuting all raising operators to the left of all $\coth$'s to resum them into one quantity. For this, one requires the definition of a reversal operator on $N$ indices, $R_N\equiv(1\,N)(2\,N-1)(3\,N-2)\cdots$ where the product continues until the middle one or two indices. For example, in terms of transpositions the first few are
\begin{align*}
	R_2=(12),\quad R_3=(13),\quad R_4=(14)(23).
\end{align*}
Since $R_N^2=1$ we can think of in terms of projection operators, $\rho_N^2=\rho_N$, where $\rho_n=\frac{1}{2}(1-(-1)^nR_n)$ with $\rho_1\equiv 1$ and $\rho_n$ projects $R_n$ onto $(-1)^{n+1}$, then 
$$(1-R_{n+1}R_n)\rho_n=(1-R_{n+1}(-1)^{n+1})\rho_n=2\rho_{n+1}\rho_n=\rho_{n+1}(1-R_{n+1}(-1)^{n+1})\rho_n.$$ 
This immediately provides the result for the final permutation quantity that we require,
\begin{align}
	P_N &=\prod_{n=1}^{N-1}(1-R_{n+1}R_n)=\prod_{n=1}^{N-1}(1-R_{n+1}R_n)\rho_1\nonumber\\
	&=\prod_{n=1}^{N-1}(2\rho_{n+1})\rho_1=\prod_{n=1}^{N-1}(1-(-1)^{n+1}R_{n+1}),\quad P_1\equiv 1\label{eq:pn-intro}
\end{align}
which provides $2^{N-1}$ permutations from a linear combination of reversal operators, and clearly $R_NP_N=(-1)^{N+1}P_N$. One can use either cyclic permutations, $R_{n+1}R_n$, which are mathematically more natural, or reversals, $R_{n+1}(-1)^{n+1}$, which are easier to think about physically but come with an alternating minus sign which must be tracked. Cyclic permutations have the action $R_NR_{N-1}\{L_1,L_2,\dots,L_{N-1},L_N\}=\{L_2,L_3,\dots,L_N,L_1\}$ which appears as though $L_1$ moves to the end of the string of variables, shifting all others left by one; this is actually the mathematical inverse of our cyclic permutation operators. Our permutation operators in fact permute the indices of the objects in the set $\{L_1,L_2,\dots,L_N\}$ and not the objects themselves.
One can formulate analogous arguments for permutations of the $B_m$'s instead as is done in Appendix~C of the previous paper in a different context.

In terms of cyclic permutations the main result is
\begin{align}
	C &= A + \sum_{N=1}^{\infty}\frac{1}{\sinh(L_1+\cdots+L_N)}P_{N}\,h_{N-1}(L_1,L_1+L_2,\dots,L_1+L_2+\cdots+L_{N-1})L_1\, B_1\cdots B_N\label{eq:resPH}.
\end{align}
Writing
\begin{align*}
	\lop H_{N-1}&=P_N\,h_{N-1} L_1\nonumber\\
	&=(1-R_NR_{N-1})P_{N-1}\,h_{N-1}L_1,\quad N>1
\end{align*}
where the second line is only valid for $N>1$, shows that half of this formula depends on $L_1$ up to $L_{N-1}$ and the other half depends on $L_2$ up to $L_N$, which is the $Q_{N-1}$ idea introduced earlier in equation~\eqref{eq:gn-rev}. Using the shorthand notation introduced in~\eqref{eq:sym}, equations~\eqref{eq:H1} to~\eqref{eq:H4} can be written as
\begin{subequations}
	\begin{align}
		H_1 = P_1\,h_1L_1 &= [L_1]\label{eq:perm1}\\
		H_2 = P_2\,h_2L_1 &= [L_1L_2]-[L_2L_1]\\
		H_3 = P_3\,h_3L_1 &= [L_1L_2L_3]\nonumber\\
		& - ([L_2L_1L_3]+[L_2L_3L_1])\nonumber\\
		&+ [L_3L_2L_1]\\
		H_4 = P_4\,h_4L_1 &=  [L_1L_2L_3L_4]\nonumber\\
		&- ([L_2L_1L_3L_4]+[L_2L_3L_1L_4]+[L_2L_3L_4L_1])\nonumber\\
		&+ ([L_3L_2L_1L_4]+[L_3L_2L_4L_1]+[L_3L_4L_2L_1])\nonumber\\
		&-[L_4L_3L_2L_1]\label{eq:perm4}
	\end{align}
\end{subequations}
then the $G_N$ can be written by using the reversal idea in equation~\eqref{eq:gn-rev} with $Q_N=H_N$. To third order in $B$ we find
\begin{align*}
	\lop H_0 &= L_1\\
	\lop H_1 &= [L_1]-[L_2]\\
	\lop H_2 &= [L_1L_2]-[L_2L_1]-[L_2L_3]+[L_3L_2].
\end{align*}
We view the permutation statement of the main result as most convenient for applications in physics, since it is a linear combination of a single hyperbolic quantity $h_N$ at each order, with arguments controlled by a sequence of reversal operators.

\subsection{Structure of the paper}
The structure of this paper is as follows.
In Section~\ref{sec:previous-paper} we summarise the relevant results from the previous paper; the original representation of $G_N$, an identity and also marching identities so this paper is more self-contained.
In Section~\ref{sec:edge-K} we develop the central idea of what we refer to as \textit{edge identities} and use them to derive a symmetric partition representation, all of this is completed at the level of generating functions, employing the identity from the previous paper in various ways. Edge identities involve incorporating an additional $h_1$ as $h_{r-1}h_1\mapsto h_r$ plus lower-order terms that we specify. The symmetric partition representation has a clean structure and we use this as our midpoint between the original and permutation representations.
In Section~\ref{sec:perm-rep} we employ the $\coth$ angle addition identity, constructing the symmetric partition representation from this. We find a recurrence relation which defines a new, permutation-based representation, ending the section with the main result, equations~\eqref{eq:resH} or equivalently~\eqref{eq:resPH}.
In Section~\ref{sec:remainder} we prove that the permutation representation generates $G_N$ by showing that the extra function vanishes as equation~\eqref{eq:xn-rev}.
In Section~\ref{sec:marching-identities} we prove that $G_N$ satisfy marching identities. The structure of the new representation allows us to straightforwardly prove these identities to all orders where they were previously only verified to tenth order. These identities were introduced in the previous paper to show the equivalence of the power series and the usual commutator representation of the BCH formula, but we find them to be important in deriving perturbation theory for this power series \cite{jones2025}.
The permutation representation can also be useful proving that $Q_N\mapsto P_N\,h_{N}L_1$ is valid in equation~\eqref{eq:gn-rev}, such that the $X_N$ difference between the original and permutation representation vanishes under reversal as stated in equation~\eqref{eq:xn-rev} and demonstrated in Appendix~\ref{app:sec-equiv}.
We also provide an overcomplete representation of $G_N$ in Appendix~\ref{app:overcomplete} that turns out to be useful in deriving part of the perturbation theory.

\section{Preliminaries}\label{sec:previous-paper}
In this section, we introduce the quantities carried over from the previous paper \cite{moodie2021}. We start with the original representation of $G_N$ and a hyperbolic identity that shows how certain sums of products of the functions $f_r$ reduce to constants. This identity lets us rewrite the original representation in a symmetric partition form, which is the starting point for our analysis. We also introduce marching identities, which are proved later.

First we require the original representation of $G_{N+1}$ and we choose to use~(5.5) to~(5.8) from the previous paper (with a factor of two extracted)
\begin{align}
	G_{N+1}(L_1,L_2,\dots,L_{N+1})&=G_1(L_1+L_2+\cdots+L_{N+1})\bigg(s_N\nonumber\\
	&\quad+\sum_{r=1}^{N}F(-L_1-L_2-\cdots-L_r,L_{r+1}+L_{r+2}+\cdots+L_{N+1})\nonumber\\
	&\quad\times f_{r-1}(-L_r,-L_{r-1},\dots,-L_2)f_{N-r}(L_{r+1},L_{r+2},\dots,L_{N})\bigg)\label{eq:gn}\\
	G_1(L_1)&=\frac{L_1}{\sinh(L_1)}\label{eq:g1}\nonumber\\
	F(x,y)&\equiv\frac{x\coth(x)-y\coth(y)}{x+y}\nonumber
\end{align}
in terms of the sequence of functions $f_r$---that are provided in the previous paper and also in equations~\eqref{eq:h0} to~\eqref{eq:h6}---and the quantity $F(x,y)$ that is just notational convenience. Since $\coth(x)\equiv f_1(x)$, this is the product of three $f_r$'s as mentioned in the introduction and the reversal structure of the right-hand side is proven in this section. The constants $s_N$ are provided in the previous paper as $a_{N+1}^\text{odd}$ and are the coefficients of the series
\begin{align}
	s(z)\equiv\frac{\sinh(z)\cosh(z)}{z} &=1+\frac{2}{3}z^2+\frac{2}{15}z^4+\frac{4}{315}z^6+\frac{2}{2835}z^8+\cdots\equiv \sum_{N=0}^{\infty}s_{N} z^{N}\equiv \sum_{N=0}^{\infty}a_{N+1}^\text{odd} z^{N}
\end{align}
where we chose to redefine the series $s_N$ to align $N$ with the other quantities in this paper.

We take the sequence of functions $f_r$ from the previous paper but redefine them in terms of $h_r$ that accumulate arguments as one tracks from left to right, for example
\begin{align}
	f_3(L_1,L_2,L_3)=h_3(L_1,L_1+L_2,L_1+L_2+L_3),\label{eq:fh}
\end{align}
which is more natural for physical applications and was discussed around equation~\eqref{eq:omit-args}. When the operator $A$ in the BCH formula is diagonal the arguments of $h_r$ are the differences between the eigenvalue of the diagonal operator in the starting state and an intermediate state; these arguments then correspond to denominators in ordinary perturbation theory. 
Including the next parameter forces us to consider $f_1(L_1+\cdots+L_r)$, which is a natural extension to $h_{r-1}(L_1,L_1+L_2,\dots,L_1+\cdots+L_{r-1})$ rather than to $f_{r-1}(L_1,L_2,\dots,L_{r-1})$.
Thinking in terms of denominators is natural here and provides the intuition for this choice, the mapping $f_2(x_1,x_2)=h_2(x_1,x_1+x_2)\mapsto\frac{1}{x_1(x_1+x_2)}$ is obvious in terms of the $h_2$ whereas $f_2$ obscures this. The denominator analogue in Appendix~\ref{app:denom} also makes this choice natural.

We take the first few $f_r$ from the previous paper and convert them into $h_r$, the first few are
\begin{subequations}
	\begin{align}
		h_0 &\equiv 1\label{eq:h0} \\
		h_1(x_1) &= \coth(x_1)\\
		h_2(x_1,x_2) &= \coth(x_1)\coth(x_2)-\frac{1}{3} \\
		h_3(x_1,x_2,x_3) &= \coth(x_1)\coth(x_2)\coth(x_3)-\frac{1}{3}(\coth(x_1)+\coth(x_3))\\
		h_4(x_1,x_2,x_3,x_4)&= \coth (x_1) \coth (x_2) \coth (x_3) \coth (x_4) \nonumber\\
		&\quad-\frac{1}{3} (\coth (x_1)\coth (x_4) +\coth (x_3)\coth
		(x_4) +\coth (x_1) \coth (x_2))+\frac{2}{15}\\
		h_5(x_1,x_2,x_3,x_4,x_5) &= \coth (x_1) \coth (x_2) \coth (x_3)\coth (x_4) \coth (x_5)\nonumber\\
		&\quad-\frac{1}{3} ( \coth (x_1)\coth (x_4) \coth
		(x_5)+\coth (x_3)\coth (x_4) \coth (x_5) \nonumber\\
		&\quad+\coth (x_1) \coth
		(x_2)\coth (x_5) +\coth (x_1) \coth (x_2) \coth (x_3))\nonumber\\
		&\quad+\frac{2}{15} (\coth
		(x_1)+\coth (x_5))+\frac{1}{9}\coth (x_3)\\
		h_6(x_1,x_2,x_3,x_4,x_5,x_6) &= \coth (x_1) \coth (x_2) \coth (x_3)\coth (x_4) \coth (x_5) \coth (x_6) \nonumber\\
		&-\frac{1}{3} (\coth (x_1)\coth(x_4) \coth (x_5) \coth (x_6) \nonumber\\
		&+\coth(x_3)\coth (x_4) \coth (x_5) \coth (x_6) +\coth (x_1) \coth (x_2) \coth (x_3)\coth (x_4)\nonumber\\
		& + \coth (x_1)	\coth (x_2)\coth (x_5) \coth (x_6)+\coth (x_1) \coth (x_2) \coth (x_3)\coth (x_6) )\nonumber\\
		&+\frac{1}{9} (
		\coth (x_1)\coth (x_4)+\coth (x_3)\coth (x_4) + \coth (x_3)\coth (x_6))\nonumber\\
		&+\frac{2}{15} (\coth (x_5)
		\coth (x_6)+\coth (x_1)\coth (x_6) +\coth (x_1) \coth (x_2))-\frac{17}{315}\label{eq:h6}
	\end{align}
\end{subequations}
that are constructed by writing down the initial product of $\coth$'s and then removing all even numbers of contiguous terms and replacing them with the appropriate coefficient of the Taylor expansion of
\begin{align}
	t(z)\equiv\frac{\tanh(z)}{z}=1 - \frac{1}{3}z^2 + \frac{2}{15}z^4 - \frac{17 }{315}z^6 + \frac{62}{2835}z^8-\frac{1382}{155925}z^{10}+\cdots\equiv\sum_{N=0}^\infty t_N z^N\label{eq:tan-over-x},
\end{align}
and this form generalises to all orders, as explained in the previous paper. Although there are several possible methods, we believe the easiest way to generate these functions is by using edge identities introduced later in equation~\eqref{eq:edge}.
These functions clearly have the property
\begin{align}
	h_N(x_1, x_2,\dots,x_N)=h_N(x_N,x_{N-1},\dots,x_1)\label{eq:hn-sym}
\end{align}
which we use in Appendix~\ref{app:overcomplete} to write these functions with the arguments in a specified order.

Next we will manipulate $G_{N+1}$ into the form required to start our investigation by isolating the terms dependent on $L_1$ up to $L_N$.
Using the form of $G_1(x)$ (equation~\eqref{eq:g1}) and multiplying by the $\sinh$, equation~\eqref{eq:gn} becomes
\begin{align}
	\sinh(L_1+L_2+\cdots+L_{N+1})&G_{N+1}(L_1,L_2,\dots,L_{N+1})=s_N(L_1+L_2+\cdots+L_{N+1})\nonumber\\
	&+\sum_{r=1}^{N}[(-L_1-L_2-\cdots-L_r)\coth(-L_1-L_2-\cdots-L_r)\nonumber\\
	&\quad-(L_{r+1}+L_{r+2}+\cdots+L_{N+1})\coth(L_{r+1}+L_{r+2}+\cdots+L_{N+1})]\nonumber\\
	&\times f_{r-1}(-L_r,-L_{r-1},\dots,-L_2)f_{N-r}(L_{r+1},L_{r+2},\dots,L_{N}).
\end{align}
If we let $r=N+1-r'$ in the second set of bracketed terms then we can recognise just one function controlling $G_{N+1}$,
\begin{align}
	\sinh(L_1+L_2+\cdots+L_{N+1})&G_{N+1}(L_1,L_2,\dots,L_{N+1})\nonumber\\
	&=E_{N}(L_1,L_2,\dots,L_{N})-(-1)^{N+1}E_{N}(L_{N+1},L_{N},\dots,L_2)
\end{align}
where the original representation is
\begin{align}
	E_{N}(L_1,L_2,\dots,L_{N})& \equiv \sum_{r=1}^{N}(-1)^{r-1}(L_1+L_2+\cdots+L_r)\nonumber\\
	&\times f_{r-1}(L_r,L_{r-1},\dots,L_2)f_1(L_1+L_2+\cdots+L_r)f_{N-r}(L_{r+1},L_{r+2},\dots,L_{N})\nonumber\\
	&+\Delta_{N}\label{eq:gn-en}\\
	\Delta_{N}&\equiv s_N\frac{1}{2}(L_1+[L_1+L_2+\cdots+L_N])\label{eq:deltan}
\end{align}
and we split the linear term between the two halves with the subtlety that $s_NL_1$ is wholly present in the first half as this variable does not appear in the reversed terms; analogously for $s_NL_{N+1}$ in the reversed half. We also used the odd/even symmetry to extract minus signs from $f_r$, and now we use $f_1(x)=\coth(x)$. From here we shall deal with only half of $G_{N+1}$ by first turning $E_N$ into a generating function and doing manipulations at that level.

The first few orders of the original representation are
\begin{subequations}
	\begin{align}
		E_0&= \Delta_0\\
		E_1(L_1) &= L_1f_1(L_1)\\
		E_2(L_1,L_2) &= L_1f_1(L_1)f_1(L_2)-(L_1+L_2)f_1(L_2)f_1(L_1+L_2)+\Delta_2\\
		E_3(L_1,L_2,L_3) &= L_1f_1(L_1)f_2(L_2,L_3)\nonumber\\
		&-(L_1+L_2)f_1(L_2)f_1(L_1+L_2)f_1(L_3)\nonumber\\
		&+(L_1+L_2+L_3)f_2(L_3,L_2)f_1(L_1+L_2+L_3)\\
		E_4(L_1,L_2,L_3,L_4) &= L_1f_1(L_1)f_3(L_2,L_3,L_4)\nonumber\\
		&-(L_1+L_2)f_1(L_2)f_1(L_1+L_2)f_2(L_3,L_4)\nonumber\\
		&+(L_1+L_2+L_3)f_2(L_3,L_2)f_1(L_1+L_2+L_3)f_1(L_4)\nonumber\\
		&-(L_1+L_2+L_3+L_4)f_3(L_4,L_3,L_2)f_1(L_1+L_2+L_3+L_4)+\Delta_4,
	\end{align}
\end{subequations}
\begin{subequations}
	\begin{align}
		\Delta_0 &= \frac{1}{2}L_1\\
		\Delta_2&=\frac{1}{2}\frac{2}{3}(L_1+[L_1+L_2])\\
		\Delta_4&=\frac{1}{2}\frac{2}{15}(L_1+[L_1+L_2+L_3+L_4])
	\end{align}
\end{subequations}
where the sum of product of three $f_r$'s is clearly visible (and in~\eqref{eq:gn-en}) and this formula is to be superseded by the new representation.
We also take the identity~(5.2) from the previous paper
\begin{align}
	s_N&=f_N(L_1,L_2,\dots,L_N)\nonumber\\
	&\quad+\sum_{r=1}^{N}(-1)^rf_{r-1}(L_r,L_{r-1},\dots,L_2)f_1(L_1+L_2+\cdots+L_r)f_{N-r}(L_{r+1},L_{r+2},\dots,L_{N})\label{eq:ml-5.2}
\end{align}
which will later be turned into generating functions in multiple ways, allowing a bridge between representations. The first term in the identity is clearly in the new representation and seeds it.

Finally we take a set of identities from the previous paper that we refer to as marching identities that we mentioned in the abstract. These are crucial to deriving perturbation theory for the BCH power series \cite{jones2025}. Two examples of marching identities are
\begin{subequations}
	\begin{align}
		&G_3(\alpha_1,\beta_1,\beta_2)+G_3(\beta_1,\alpha_1,\beta_2)+G_3(\beta_1,\beta_2,\alpha_1)=0\label{eq:g3}\\
		&G_6(\alpha_1,\alpha_2,\beta_1,\beta_2,\beta_3,\beta_4)+G_6(\alpha_1,\beta_1,\alpha_2,\beta_2,\beta_3,\beta_4)+G_6(\alpha_1,\beta_1,\beta_2,\alpha_2,\beta_3,\beta_4)\nonumber\\	
		&\quad+G_6(\alpha_1,\beta_1,\beta_2,\beta_3,\alpha_2,\beta_4)+G_6(\alpha_1,\beta_1,\beta_2,\beta_3,\beta_4,\alpha_2)+G_6(\beta_1,\alpha_1,\alpha_2,\beta_2,\beta_3,\beta_4)\nonumber\\	
		&\quad+G_6(\beta_1,\alpha_1,\beta_2,\alpha_2,\beta_3,\beta_4)+G_6(\beta_1,\alpha_1,\beta_2,\beta_3,\alpha_2,\beta_4)+G_6(\beta_1,\alpha_1,\beta_2,\beta_3,\beta_4,\alpha_2)\nonumber\\	
		&\quad+G_6(\beta_1,\beta_2,\alpha_1,\alpha_2,\beta_3,\beta_4)+G_6(\beta_1,\beta_2,\alpha_1,\beta_3,\alpha_2,\beta_4)+G_6(\beta_1,\beta_2,\alpha_1,\beta_3,\beta_4,\alpha_2)\nonumber\\	
		&\quad+G_6(\beta_1,\beta_2,\beta_3,\alpha_1,\alpha_2,\beta_4)+G_6(\beta_1,\beta_2,\beta_3,\alpha_1,\beta_4,\alpha_2)+G_6(\beta_1,\beta_2,\beta_3,\beta_4,\alpha_1,\alpha_2)=0\label{eq:g6}
	\end{align}
\end{subequations}
where the arguments $\alpha_i$ and $\beta_j$ may take any values. We refer to these as marching identities
\footnote{Marching identities have the structure of \textit{shuffle products} which have been studied in relation to free Lie algebras and the BCH formula \cite{Reutenauer1993}. For example in terms of a shuffle product, $\shuffle$, the two marching identities in equations~\eqref{eq:g3} and~\eqref{eq:g6} are
	\begin{align*}
		\alpha_1\shuffle\beta_1\beta_2 &=\alpha_1 \beta_1 \beta_2 +\beta_1 \alpha_1 \beta_2 +\beta_1 \beta_2 \alpha_1 =0 \\
		\alpha_1\alpha_2\shuffle\beta_1\beta_2\beta_3\beta_4 &=\alpha_1 \alpha_2 \beta_1 \beta_2 \beta_3 \beta_4 +\alpha_1 \beta_1 \alpha_2 \beta_2 \beta_3 \beta_4 +\alpha_1 \beta_1 \beta_2 \alpha_2 \beta_3 \beta_4 +\alpha_1 \beta_1 \beta_2 \beta_3 \alpha_2 \beta_4 +\alpha_1 \beta_1 \beta_2 \beta_3 \beta_4 \alpha_2\nonumber 	\\
		&+\beta_1 \alpha_1 \alpha_2 \beta_2 \beta_3 \beta_4 +\beta_1 \alpha_1 \beta_2 \alpha_2 \beta_3 \beta_4 +\beta_1 \alpha_1 \beta_2 \beta_3 \alpha_2 \beta_4 +\beta_1 \alpha_1 \beta_2 \beta_3 \beta_4 \alpha_2 +\beta_1 \beta_2 \alpha_1 \alpha_2 \beta_3 \beta_4 \nonumber\alpha_2 \beta_3 \nonumber\\
		&+\beta_1 \beta_2 \alpha_1 \beta_3 \alpha_2 \beta_4 +\beta_1 \beta_2 \alpha_1 \beta_3 \beta_4 \alpha_2 +\beta_1 \beta_2 \beta_3 \alpha_1 \alpha_2 \beta_4 +\beta_1 \beta_2 \beta_3 \alpha_1 \beta_4 \alpha_2 +\beta_1 \beta_2 \beta_3 \beta_4 \alpha_1 \alpha_2 =0
	\end{align*} where the concatenation of the arguments is understood to be a $G_N$ of that order, for example $\alpha_1\beta_1\alpha_2\beta_2\beta_3\beta_4\equiv G_6(\alpha_1,\beta_1,\alpha_2,\beta_2,\beta_3,\beta_4)$.}
because in equation~\eqref{eq:g3} $\alpha_1$ \textit{marches} through the other two variables and in equation~\eqref{eq:g6} one block of parameters, $\alpha_1$ and $\alpha_2$, marches through another block, $\beta_1$, $\beta_2$, $\beta_3$ and $\beta_4$, maintaining the relative ordering in each block. One can construct the latter identity by putting $\alpha_1\alpha_2$ at the start and pushing $\alpha_2$ through, then moving $\alpha_1\alpha_2$ to the right by one and pushing $\alpha_2$ through from there and repeat until $\alpha_1\alpha_2$ is at the right edge. The first and last terms contain the separated blocks and all other terms contain mixed blocks of parameters. We use different letters for the blocks here to make the identities clearer. There is an identity for marching any number of variables through any other number of variables and we formulate these identities using permutation operators during the proof later.

This concludes the preliminaries so that we have all that we require from the previous paper and all further work is new. Next we introduce what we call edge identities as a means to find a representation with symmetric partition structure.

\section{Manipulation of the original representation}\label{sec:edge-K}
In this section we formulate the procedure for reducing the number of $h_r$ factors in expressions when they are of the form $h_{r-1}h_1$ by finding what we refer to as \textit{edge identities}. We derive edge identities by first writing $h_r$ in terms of $h_{r-1}h_1$ and lower-order terms that are controlled by a generating function. Edge identities are the inverse of this, allowing one to incorporate an $h_{r-1}$ and an $h_1$, $h_{r-1}h_1\mapsto h_r$ and lower-order terms. We then construct edge identities for functions with reversed arguments and also convert the identity~\eqref{eq:ml-5.2} into generating functions that allows simplifications later. Finally in this section, we use the results to transform the original representation into the symmetric partition representation.
This leaves a linear combination of just two $h_r$'s rather than three as suggested in the introduction, which we can target in the next section from a simple hyperbolic identity. 

We encounter $\sop$ as an argument and therefore ordering is important. This allows us to extract shifts so that all quantities have arguments starting at $L_1$. It is crucial for us that all terms in each generating function have the same form, that is, starting from $L_1$ and increasing up to $L_N$ at order $z^N$ in the generating function (or the arguments in reverse from $L_N$ down to $L_1$). This enables us to recognise convolutions and separate multiple sums into products of generating functions.
We provide an example of extracting the shift by rewriting equation~\eqref{eq:ml-5.2},
\begin{align}
	s_N&=\sum_{r=0}^{N}(-1)^r\tilde u_r\sop^r h_{N-r}\sop^{-r}\label{eq:ml-5.2-rewritten}\\
	\tilde u_r&\equiv h_{r-1}(L_{r-1},L_{r-1}+L_{r-2},\dots,L_2+\cdots+L_{r-1})h_1(L_1+\cdots+L_r),\quad \tilde u_0\equiv 1\nonumber
\end{align}
where we use $h_r$ instead of $f_r$ and we have introduced $\tilde u_r$ to describe terms with reversed arguments and also incorporate the $r=0$ term into the sum. We extracted $\sop^r$ by using $h_{N-r}(L_{r+1},\dots,L_{r+1}+\cdots+L_{N})=\sop^r h_{N-r}(L_{1},\dots,L_1+\cdots+L_{N-r})\sop^{-r}\equiv \sop^r h_{N-r}\sop^{-r}$ so that all functions with omitted arguments start from $L_1$. 
When we omit arguments from a reversed function indexed by $r$, such as $\tilde u_r$, the first argument is $L_r$ and subsequent arguments decrease to $L_1$. For example, the above can be written as
\begin{align*}
	\tilde u_r&=h_{r-1}(L_{r-1},\dots,L_2+\cdots+L_{r-1})h_1(L_1+\cdots+L_r)\\
	&=f_{r-1}(L_{r-1},\dots,L_2)f_1(L_1+\cdots+L_r)\\
	&=\sop \tilde h_{r-1}\sop^{-1}h_1(L_1+\cdots+L_r) = \sop \tilde f_{r-1}\sop^{-1}f_1(L_1+\cdots+L_r).
\end{align*}
We use $\tilde\cdot$ to represent quantities with reversed arguments throughout this document. 
Extracting $\sop^p$ from reversed quantities such as $h_r(L_{p+r},L_{p+r-1},\dots,L_{p+1})$, allows us to write all functions indexed by $p$ as arguments of $L_r$ down to $L_1$, in this example $\sop^p h_r(L_{r},L_{r-1},\dots,L_{1})\sop^{-p}=\sop^p\tilde h_r\sop^{-p}$. This makes the quantities susceptible to generating functions while having the added benefit of being able to omit arguments.
Note that $\sop^p$ is also required during partitioning arguments as the second term in a partitioned term starts at $L_{p+1}$ and we write this as $\sop^pL_1\sop^{-p}$ and so on to start at $L_1$ and to make a physical number instead of leaving an operator.
We refer to the final expression on the first line as a \textit{number-convolution} to distinguish it from an operator convolution because one quantity operates with $\sop^p$ on the other quantity and terminates the raising operation with $\sop^{-p}$ so that the effect is local and does not produce an operator unless the other quantity itself is an operator. Number-convolutions are not susceptible to generating function analysis, but nevertheless can sometimes be inverted in an analogous way as in Appendix~\ref{app:number-gen}.

Where possible, we tend to use the summation index $m$ when the summands are $\tilde h_m$ and $h_{m}$, and $r,p,\dots$ when sums involve one or two $\tilde u_r$'s or $u_r$'s. This makes the distinction clearer between sums over three or two $h_m$'s.

\subsection{Edge identities}
We convert identity~\eqref{eq:ml-5.2-rewritten} into generating functions and this will provide a crucial relationship for finding the symmetric partition representation. To do this we first require edge identities, which are constructed by looking at the right edge of the product of $\coth$'s in $h_m$ and appreciating that it starts with either a $\coth$ or with a gap and this allows us to write down the relationship
\begin{equation}\label{eq:edge}
	\begin{aligned}
		h_m(x_1,&x_1+x_2,\dots,x_1+x_2+\cdots+x_m) =\nonumber\\
		&\sum_{r=0}^{m-1}h_{m-r-1}(x_1,x_1+x_2,\dots,x_1+x_2+\cdots+x_{m-r-1})h_1(x_1+x_2+\cdots+x_{m-r})t_r+t_m\\
		&\equiv\sum_{r=0}^{m}t_r u_{m-r},\quad u_0\equiv 1
	\end{aligned}
\end{equation}
for a set of parameters $\{x_1,x_2,\dots,x_m\}$ where $x_n\mapsto L_n$ provides edge identities later and $x_n\mapsto L_{m+1-n}$ provides the reversed analogue in the next subsection, and $t_n$ are the coefficients in the Taylor expansion of $\tanh(z)/z$ given in equation~\eqref{eq:tan-over-x}. The $h_1$ guarantees the gap is over and the $h_{m-q-1}$ allows arbitrary subsequent gaps, where the gap is of size $q$. We can then consider two generating functions
\begin{align*}
	h(z) &= \sum_{m=0}^\infty h_m(x_1,x_1+x_2,\dots,x_1+x_2+\cdots+x_m)z^m\equiv \sum_{m=0}^{\infty}h_mz^m\\
	u(z) &= 1 + \sum_{r=1}^\infty h_{r-1}(x_1,x_1+x_2,\dots,x_1+x_2+\cdots+x_{r-1})h_1(x_1+x_2+\cdots +x_r)z^r\equiv\sum_{r=0}^\infty u_rz^r
\end{align*}
and the previous relationship provides
\begin{align*}
	h(z) &= \sum_{m=0}^{\infty}\sum_{r=0}^m t_ru_{m-r}z^n=\sum_{m=0}^\infty\sum_{r=0}^{m}t_ru_{m-r}z^{m-r}z^r\\
	&=\sum_{r=0}^\infty t_rz^r\sum_{m=r}^\infty u_{m-r}z^{m-r}=t(z)u(z)
\end{align*}
from which we deduce the reciprocal relationship
\begin{align}
	u(z)=T(z)h(z)
\end{align}
where $T(z)\equiv 1/t(z)$ generates the following coefficients
\begin{align}
	T(z)\equiv\frac{z}{\tanh(z)}=1+\frac{1}{3}z^2-\frac{1}{45}z^4+\frac{2}{945}z^6-\frac{1}{4725}z^8+\frac{2}{93555}z^{10}\cdots\equiv\sum_{p=0}^\infty T_pz^p.
\end{align}
The reciprocal relationship generates what we refer to as edge identities 
\begin{align}
	&h_{r-1}(x_1,x_1+x_2,\dots,x_1+x_2+\cdots+x_{r-1})h_1(x_1+x_2+\cdots+x_r) =\nonumber\\
	&\qquad \sum_{p=0}^r T_p h_{r-p}(x_1,x_1+x_2,\dots,x_1+x_2+\cdots+x_{r-p})\nonumber\\
	&u_r=\sum_{p=0}^r T_p h_{r-p}\label{eq:redge}
\end{align}
where the final line is simply a rewriting in our shorthand notation. Edge identities are absolutely crucial in the later analysis, allowing us to rewrite the product of two $h_m$'s as one, so that the original representation can be written in terms of two $h_m$'s instead of three as suggested in the introduction. We will see that this idea allows us to find a symmetric partition representation that we can relate to the new permutation representation in Section~\ref{sec:perm-rep}.

\subsection{Writing an identity in terms of generating functions}
In this section we use edge identities on identity~\eqref{eq:ml-5.2-rewritten} to convert it into generating functions and these results will be drawn upon later.
The fact that we have number-convolutions makes this section long-winded. We cannot use generating functions so the derivation is for each $N$ individually. We show that one of the quantities introduced later $W(z)$ actually just generated numbers and not operators so that number-convolution collapses and generating functions can be employed again.
We require the reversed analogue of edge identities so first define
\begin{align*}
	\tilde h(\sop z) &\equiv \sum_{m=0}^\infty h_m(L_m,\dots,L_1+\cdots+L_m)(-\sop z)^m\equiv\sum_{m=0}^{\infty}\tilde h_m(-\sop z)^m
\end{align*}
then we can write edge identities for reversed arguments
\begin{align}
	\tilde u(\sop z) &\equiv 1 + \sum_{r=1}^\infty h_{r-1}(L_r,\dots,L_{2}+\cdots+L_r)h_1(L_1+\cdots L_r)(-\sop z)^r\equiv\sum_{r=0}^\infty \tilde u_r(-\sop z)^r,\nonumber\\
	\tilde u(\sop z)&= T(\sop z)\tilde h(\sop z)\nonumber\\
	\tilde u_q(-\sop)^{q} &=\sum_{p=0}^{q}T_p\sop^p\tilde h_{q-p}(-\sop)^{q-p}\label{eq:reversed-edge}
\end{align}
where the final line is the analogue of~\eqref{eq:redge}. We extracted $\sop^m$ as explained above to have functions of reversed arguments ending at $L_1$ so as to be susceptible to generating functions.

There are multiple places in this document where we rewrite a sum over $r$ of $u_r$ (and/or $\tilde u_r$) as a sum over $m$ and $r$ of $h_{r-m}$ ($\tilde h_{r-m}$). Therefore, we specify the generic procedure once in detail and refer back to this when it is employed:
\begin{enumerate}
	\item It is assumed that the starting point is a sum over $r$ of $u_r$ (and/or $\tilde u_r$). 
	\item Substitute in the edge-identity~\eqref{eq:redge} (reversed edge-identity~\eqref{eq:reversed-edge}) to eliminate $u_r$ ($\tilde u_r$) in favour of a sum over $m$ of $h_{r-m}$ ($\tilde h_{r-m}$).
	\item Swap the order of summation to make the sum over $r$ internal, thereby shifting the $r$ sum to start from $m$.
	\item Shift the now-internal sum over $r$ down to zero to recognise a previously defined or targeted quantity as a sum over $h_{r-m}$'s ($\tilde h_{r-m}$'s).
\end{enumerate}

We now employ the procedure for the first time for equation~\eqref{eq:ml-5.2-rewritten}. First use the reversed edge identity given in equation~\eqref{eq:reversed-edge} to rewrite $\tilde u_r$ as the sum over single $\tilde h_{r-m}$'s, and substitute into equation~\eqref{eq:ml-5.2-rewritten} to find
\begin{align*}
	s_N &= \sum_{r=0}^{N}\left(\sum_{p=0}^{r}T_p\sop^p\tilde h_{r-p}\sop^{-p}\right)(-\sop)^{r}h_{N-r}\sop^{-r}.
\end{align*}
Then use that only even $p$ terms exist to write $\sop^{-p}$ as $(-\sop)^{-p}$, and swap the order of summation,
\begin{align*}
	s_N &= \sum_{p=0}^{N}T_p\sop^p\sum_{r=p}^{N}\tilde h_{r-p}(-\sop)^{r-p}h_{N-r}\sop^{-r},
\end{align*}
and finally shift the sum over $r$ to start from $0$ using $r-p=q$,
\begin{align}
	s_N &=\sum_{p=0}^{N}T_p\sop^p\sum_{q=0}^{N-p}\tilde h_q(-\sop)^qh_{N-p-q}\sop^{-q}\sop^{-p}\equiv\sum_{p=0}^{N}T_p\sop^pW_{N-p}\sop^{-p}\label{eq:id-num}\\
	W_N &\equiv\sum_{m=0}^{N}\tilde h_m(-\sop)^m h_{N-m}\sop^{-m}\nonumber
\end{align}
where $W_N$ is the analogue of $s_N$ when there are only two $h_m$'s.
Inverting the first line using the procedure outlined in Appendix~\ref{app:number-gen} for number-convolutions provides
\begin{align*}
	W_N &= \sum_{m=0}^{N}t_m\sop^ms_{N-m}\sop^{-m}=\sum_{m=0}^{N}t_ms_{N-m}
\end{align*}
and so $W_N$ non-trivially inherits being a number from the fact that $s_{N-m}$ is just a number, and is generated by
\begin{align*}
	W(z) &= t(z)s(z)=\left(\frac{\sinh z}{z}\right)^2.
\end{align*}
We will use the inverse relationship
\begin{align*}
	s(z)&=T(z)W(z)\\
	s_N&=\sum_{p=0}^{N}T_pW_{N-p}
\end{align*}
several times when we cancel terms arising from the identity with terms occurring in the new representation and by eradicating $s_N$ in favour of $W_N$ the arguments are clearer.
The resulting expressions are simpler because $W_N$ is just a number (stemming from equation~\eqref{eq:id-num}), and we use this later in deriving the symmetric partition representation, equations~\eqref{eq:vjk} and~\eqref{eq:vkk}.

\subsection{Symmetric partition representation}
Using edge identities and the relations derived from the previous paper's identity we can now find the symmetric partition representation. Our strategy is simply to convert $E_N$ into a generating function, using a variety of identities from the previous section to eventually recognise a symmetric partition representation which sets up the derivation of the permutation representation in the next section.

First we define the generating functions
\begin{align*}
	E(z) &\equiv \sum_{N=0}^{\infty} E_{N}(L_1,L_2,\dots,L_{N})z^N,\\
	\Delta(z) &\equiv \sum_{N=0}^{\infty}\Delta_N z^N
\end{align*}
where $E_N$ and $\Delta_N$ are defined in equations~\eqref{eq:gn-en} and~\eqref{eq:deltan} and we rewrite $E_N$ as
\begin{align*}
	E_N &= -\sum_{r=0}^{N}\tilde u_r (-\sop)^r h_{N-r}\sop^{-r}(L_1+\cdots+L_r)+\Delta_N.
\end{align*}

The first step is to use the geometric series result $L_1+\cdots+L_r=(1-\sop^r)(1-\sop)^{-1}L_1\equiv(1-\sop^r)H_0$, where $H_0$ is singular and unphysical but at the level of $G_N$ things remain regular because one deals with $\lop H_0=L_1$. Introducing $H_0$ and separating it from $1-\sop^r$ is subtle---it allows us to factorise hyperbolic quantities from the $L_1+\cdots+L_r$ at the generating function level. Indeed, in the next section we derive a purely hyperbolic quantity and connect back to this section by multiplying the result by $H_0$.
We also use the procedure from the previous section that allowed us to reach~\eqref{eq:id-num}. This allows us to convert $\tilde u_r$ into a sum over $\tilde h_{r-p}$, swap the order of summation and finally shift the now-internal sum over $r$ down to start at $0$, providing
\begin{align*}
	E_N &= -\sum_{p=0}^{N}T_p\sop^p\sum_{m=0}^{N-p}\tilde h_m(-\sop)^mh_{N-p-m}\sop^{-p}\sop^{-m}(1-\sop^{p+m})H_0+\Delta_N\\
	&\equiv(Y^{p+m}_N-Y^0_N)H_0+\Delta_N\\
\end{align*}
where we defined the central quantity in this analysis,
\begin{align*}
	Y_N^\alpha &\equiv \sum_{p=0}^{N}T_p\sop^p\sum_{m=0}^{N-p}\tilde h_m(-\sop)^mh_{N-p-m}\sop^{-p}\sop^{-m}\sop^{\alpha}
\end{align*}
which we shall consider for $\alpha\in\{0,p,p+m,N\}$. We have encountered the $\alpha=0$ case in equation~\eqref{eq:id-num} which is two number-convolutions and the identity from the previous paper, $Y_N^0=s_N$. The next two cases are non-trivial. Firstly $\alpha=p$ provides
\begin{align*}
	Y_N^p &= \sum_{p=0}^{N}T_p\sop^p\sum_{m=0}^{N-p}\tilde h_m(-\sop)^mh_{N-p-m}\sop^{-m}=\sum_{p=0}^{N}T_p\sop^pW_{N-p}
\end{align*}
which can be generated by $Y^p(z)=T(\sop z)W(z)$ which is an operator but has no $L_m$ dependence, and the unpleasant number-convolutions were dealt with in the previous section, since $W_N$ is in fact just a number.

The next case of $\alpha=p+m$ provides
\begin{align*}
	Y_N^{p+m} &=\sum_{p=0}^{N}T_p\sop^p\sum_{m=0}^{N-p}\tilde h_m(-\sop)^mh_{N-p-m}\equiv \sum_{p=0}^{N}T_p\sop^pV_{N-p}
\end{align*}
where we defined the \textit{hyperbolic core} of the problem as
\begin{align}
	V_N\equiv\sum_{m=0}^{N}\tilde h_m(-\sop)^mh_{N-m}\label{eq:vn},
\end{align}
which will be the target of the next section.
The hyperbolic core is generated by
\begin{align*}
	V(z)\equiv\tilde h(\sop z)h(z)
\end{align*}
such that $Y^{p+m}(z)=T(\sop z)V(z)$.

The fourth case, $\alpha=N$, provides an operator version of the identity in equation~\eqref{eq:id-num}, $Y_N^N=s_N\sop^N$, and is generated by $Y^N(z)=s(\sop z)=T(\sop z)W(\sop z)$. Inserting the results into the original representation provides
\begin{align}
	E(z) &= T(\sop z)V(z)H_0-s(z)H_0+\Delta(z)\label{eq:ev}.
\end{align}
where $V(z)H_0$ is what we alluded to as the symmetric partition representation. Crucially $V(z)$ is purely hyperbolic with no linear factors, meaning we can target this and construct it from the $\coth$ angle addition formula then connect it to this analysis by applying the result to $H_0$.

This completes the main goal of the section; to find a symmetric partition representation in terms of just two $h_r$ factors instead of three. We have the required formulae to derive the new representation---a linear combination of a single $h_r$---in Section~\ref{sec:perm-rep}.
However, before proceeding we provide some results that we draw upon in Section~\ref{sec:remainder}.
We find two other representations which each have partition structure. In Section~\ref{sec:remainder} we use the average of these representations in proving that the permutation representation is equivalent to the original at the level of $G_N$.
We find the other representations by considering the two other possible differences $Y_N^{p+m}-Y_N^p$ and $Y_N^{p+m}-Y_N^N$. This will provide different combinations of $h_r$'s which are compensated with a different linear term. We find
\begin{align}
	(Y_N^{p+m}-Y_N^p)H_0 &=-\sum_{p=0}^{N}T_p\sop^p\sum_{m=0}^{N-p}\tilde h_m(-\sop)^mh_{N-p-m}\sop^{-m}\sop^{-p}(L_{1+p}+\cdots+L_{p+m})\nonumber\\
	&\equiv\sum_{p=0}^{N}T_p\sop^pJ_{N-p}\sop^{-p}\nonumber\\
	J_{N}&\equiv-\sum_{m=0}^{N}(L_{1}+\cdots+L_{m})\tilde h_m(-\sop)^mh_{N-m}\sop^{-m}\label{eq:k}
\end{align}
which defines $J_N$, and
\begin{align}
	(Y_N^{p+m}-Y_N^N)H_0 	&=\sum_{p=0}^{N}T_p\sop^p\sum_{m=0}^{N-p}\tilde h_m(-\sop)^mh_{N-p-m}\sop^{-m}\sop^{-p}(L_{1+p+m}+\cdots+L_{N})\nonumber\\
	&\equiv\sum_{p=0}^{N}T_p\sop^pK_{N-p}\sop^{-p}\nonumber\\
	K_{N}&\equiv\sum_{m=0}^{N}\tilde h_m(-\sop)^m(L_{1}+\cdots+L_{N-m})h_{N-m}\sop^{-m}\label{eq:k-tilde}
\end{align}
which defines $K_N$.
In terms of generating functions we can then write
\begin{align*}
	\left[Y^{p+m}(z)-Y^p(z)\right]H_0&=T(\sop z)\left[V(z)-W(z)\right]H_0\equiv T(\sop z)K(z)\\
	\left[Y^{p+m}(z)-Y^N(z)\right]H_0&=T(\sop z)\left[V(z)-W(\sop z)\right]H_0\equiv T(\sop z)J(z)
\end{align*}
where $J(z)$ and $K(z)$ are the alternative hyperbolic quantities one may use to represent $E(z)$ and generate the quantities in equations~\eqref{eq:k} and~\eqref{eq:k-tilde} respectively. Isolating $V(z)$ provides the relations
\begin{align}
	V(z)H_0&=J(z)+W(z)H_0=K(z)+W(\sop z)H_0\label{eq:vjk}\\
	V(z)H_0&=\frac{1}{2}\left[J(z)+K(z)+(W(z)+W(\sop z))H_0\right]\label{eq:vkk}
\end{align}
In the next section we construct $V(z)H_0$ by using the angle addition identity for $\coth$ and multiplying the result by $H_0$ to connect to equation~\eqref{eq:ev}. We find that $V(z)H_0$ is proportional to a new, permutation representation.

\section{Permutation representation}\label{sec:perm-rep}
In this section we relate the original representation to one that is just a linear combination of a single $h_N$, continuing the progression of the paper that saw formulae containing three of these functions reduce to two in the previous section. We start with a basic hyperbolic identity and show that one can derive the partition formula $V(z)$ from this, finding that it satisfies a recursion relation which generates a permutation based representation. We do this partly through generating functions and also by manipulating finite sums into appropriate forms. This produces a solution as a product of $\lop$ operators and individual $\coth$ functions. We then relate this to the original representation through equation~\eqref{eq:ev}. Finally in this section we observe that the permutation formula can be written as a linear combination of reversal operators applied to one $h_r$-function at each order.
We use the initial form of the permutation representation to prove the equivalence of representations in the next section. The latter form is used in Appendix~\ref{app:sec-equiv} in a second proof of equivalence, and also in Section~\ref{sec:marching-identities} to prove that marching identities are obeyed.
A denominator analogue of this derivation is provided in Appendix~\ref{app:denom} as an instructive example which shows how permutations emerge from the recursive use of denominator identities.

The hyperbolic identity at the core of this derivation is the angle-addition formula for $\coth$,
\begin{align*}
	\coth(a+b) &= \frac{\coth(a)\coth(b)+1}{\coth(a)+\coth(b)}
\end{align*}
and we multiply by the denominator on both sides to reach our starting point,
\begin{align*}
	\coth(a)\coth(b)+1 &= \coth(a+b)\left[\coth(a)+\coth(b)\right].
\end{align*}
Next we introduce $a=L_1+\cdots+L_r$ and $b=L_{r+1}+\cdots+L_N=\sop^r (L_1+\cdots+L_{N-r})\sop^{-r}$, providing
\begin{align}
	[c_r \sop^r c_{N-r} \sop^{-r}]+1&= c_N \left[c_r+\sop^r c_{N-r}\sop^{-r}\right]\label{eq:start-coth}
\end{align}
where we also introduced the shorthand use $c_r\equiv \coth(L_1+\cdots+L_r)$ which is useful since all individual $\coth$'s start from $L_1$ and contain the sum up to $L_r$; thus requiring just one index. To construct a quantity analogous to the partition formula $V(z)$ we left-multiply~\eqref{eq:start-coth} by
\begin{align*}
	h_{r-1}(L_r,\dots,L_2+\cdots+L_{r-1})=\sop \tilde h_{r-1} \sop^{-1}
\end{align*}
and right-multiply by 
\begin{align*}
	h_{N-1-r}(L_{r+1},\dots,L_{r+1}+\cdots+L_N)\sop^r=\sop^{r} h_{N-1-r}
\end{align*}
which provides an operator and allows generating functions,
such that equation~\eqref{eq:start-coth} becomes
\begin{align*}
	\tilde r_m\sop^m r_{N-m}+\sop\tilde h_{m-1}\sop^{-1}\sop^mh_{N-1-m}=c_N\left[\tilde r_m\sop^m h_{N-1-m}+\sop\tilde h_{m-1}\sop^{-1}\sop^m r_{N-m}\right].
\end{align*}
We then insert $(-1)^r$ and sum over $r$ to provide
\begin{align}
	&\sum_{r=1}^{N-1}\tilde u_r(-\sop)^r u_{N-r}+\sop\tilde h_{r-1}\sop^{-1}(-\sop)^rh_{N-1-r}\\
	&\quad=c_N\left[\sum_{r=1}^{N}\tilde u_r(-\sop)^r h_{N-1-r}+\sop\tilde h_{r-1}\sop^{-1}(-\sop)^r u_{N-r}\right]\label{eq:checkpoint}.
\end{align}
Our goal is to find each of the four sums in terms of $h_r$ and $\tilde h_r$ with sums starting from $r=0$. This will allow us to recognise the symmetric partition representation $V_N$ defined in~\eqref{eq:vn}. The exterior $c_N$ provides the next order of the recurrence relation, which we solve to find the permutation relation.
The current expressions involve $u_r$ and $\tilde u_r$, and start from $r=1$. We recall definitions $u_0\equiv 1$ and $\tilde u_0\equiv 1$ for the following analysis.
Our strategy is to include the missing terms by adding and subtracting them in each sum, after which we apply the procedure outlined prior to equation~\eqref{eq:id-num} to recast the sums in terms of $h_r$ and $\tilde{h}_r$. This leads to the identification of $V_N$ and a recurrence relation controlled by the exterior $c_N$ on the right-hand side of~\eqref{eq:checkpoint} whose solution yields the desired permutation representation.

Firstly, the first term of the left-hand side of~\eqref{eq:checkpoint} requires including the $r=0$ and $r=N$ into the sum terms then subtracting them,
\begin{align}
	\sum_{r=1}^{N-1}\tilde u_r(-\sop)^r u_{N-r}=\sum_{r=0}^{N}\tilde u_r(-\sop)^r u_{N-r}-u_N-\tilde u_N(-\sop)^N\label{eq:rr}.
\end{align}
The second sum on the left-hand side of~\eqref{eq:checkpoint} simply requires shifting the summation index down,
\begin{align*}
	\sum_{r=1}^{N-1}\sop\tilde h_{r-1}\sop^{-1}(-\sop)^rh_{N-1-r}=-\sop\sum_{r=0}^{N-2}\tilde h_{r}(-\sop)^rh_{N-2-r}=-\sop V_{N-2}
\end{align*}
which we recognised from~\eqref{eq:vn} as being directly what we require. Then for the right-hand side of~\eqref{eq:checkpoint} we deal with the bracketed terms. The first term requires including the $r=0$ term and the second term requires shifting the index down then including the $r=N-1$ term, providing
\begin{align*}
	\sum_{r=1}^{N-1}\tilde u_r(-\sop)^r h_{N-1-r}&=\sum_{r=0}^{N-1}\tilde u_r(-\sop)^r h_{N-1-r}-h_{N-1}\\
	\sum_{r=1}^{N-1}\sop\tilde h_{r-1}\sop^{-1}(-\sop)^r u_{N-r}&=-\sop\sum_{r=0}^{N-1}\tilde h_r(-\sop)^r u_{N-1-r}+\sop\tilde h_{N-1}(-\sop)^{N-1}
\end{align*}
respectively.
The final term on each line cancels with the two final terms in equation~\eqref{eq:rr} when one uses $h_{N-1}c_N=u_N$ and $c_N\sop\tilde h_{N-1}=\tilde u_N$.
Equation~\eqref{eq:checkpoint} is now in the form
\begin{align}
	&\sum_{r=0}^{N}\tilde u_r(-\sop)^r u_{N-r}-\sop V_{N-2}=c_N\left[\sum_{r=0}^{N-1}\tilde u_r(-\sop)^r h_{N-1-r}-\sop\tilde h_{r}(-\sop)^r u_{N-1-r}\right]\label{eq:checkpoint-2}
\end{align}
and the next round of algebra requires turning each $u_r$ into $h_r$'s to recognise the partition formula. We do not need to manipulate the second term on the left-hand side as this is already in the required form.
Each term on the right-hand side requires applying the procedure used to obtain equation~\eqref{eq:id-num}, but now to $\tilde u_r$ and $u_{N-r}$ separately.
The first term on the left-hand side is our first encounter with $\tilde u_r$ and $u_{N-r}$ under a single sum. For this we can still use the procedure but it is complicated by the fact that we require both forward and reversed edge identities for one expression.
We deal with this latter case first: employing edge identities on both terms, then swapping the order of summation and relabelling the innermost sum as follows,
\begin{align}
	\sum_{r=0}^{N}\tilde u_r(-\sop)^r u_{N-r} &=\sum_{r=0}^{N}\left(\sum_{p=0}^{r}T_p\sop^p\tilde h_{r-p}\sop^{-p}\right)(-\sop)^r \left(\sum_{q=0}^{N-r}T_qh_{N-q-r}\right)\nonumber\\
	&=\sum_{p=0}^{N}T_p\sop^p\sum_{q=0}^{N-p}T_q\sum_{r=p}^{N-q}\tilde h_{r-p}(-\sop)^{r-p}h_{N-q-r}\nonumber\\
	&=\sum_{p=0}^{N}T_p\sop^p\sum_{q=0}^{N-p}T_q\sum_{r=0}^{N-p-q}\tilde h_{r}(-\sop)^{r}h_{N-p-q-r}\nonumber\\
	&=\sum_{p=0}^{N}T_p\sop^p\sum_{q=0}^{N-p}T_qV_{N-p-q}\label{eq:lhs-1}.
\end{align}
The right-hand side is more straightforward and only requires one edge identity on each term before we can reorder the sums, relabel the innermost one and recognise $V$,
\begin{align}
	\sum_{r=0}^{N-1}\tilde u_r(-\sop)^r h_{N-1-r}&=\sum_{p=0}^{N-1}T_p\sop^pV_{N-1-p}\label{eq:rhs-1}\\
	-\sop\sum_{r=0}^{N-1}\tilde h_r(-\sop)^r u_{N-1-r} &=-\sop\sum_{p=0}^{N-1}T_pV_{N-1-p}.\label{eq:rhs-2}
\end{align}
This completes the first stage of the proof, having turned each of the four terms into $V_N$. 
We combine the results in equations~\eqref{eq:lhs-1},~\eqref{eq:rhs-1},~\eqref{eq:rhs-2} to rewrite equation~\eqref{eq:checkpoint-2}, providing a recurrence relation for $V_N$,
\begin{align}
	\sum_{p=0}^{N}T_p\sop^p\sum_{q=0}^{N-p}T_q V_{N-p-q}-\sop V_{N-2}&=c_N\sum_{p=0}^{N-1}T_p(\sop^p-\sop)V_{N-1-p}\nonumber\\
	&=c_N\lop\sum_{p=0}^{N-1}T_p\frac{\sop^p-\sop}{1-\sop}V_{N-1-p}\label{eq:v-rec}
\end{align}
where in the second line we inserted $1=(1-\sop)/(1-\sop)=\lop(1-\sop)^{-1}$ left of $V_{N-1-p}$. 
We now focus on the recurrence relation controlled by $c_N\lop$ and solve this using generating functions.

To write~\eqref{eq:v-rec} as generating functions and recognise how the symmetric partition representation relates to the new representation we first apply~\eqref{eq:v-rec} to $H_0$ in order to connect to equation~\eqref{eq:ev} of the previous section. The right-hand side is then defined to be the transient, generatable $I_N$,
\begin{align*}
	I_N\equiv c_N\lop H_{N-1},\quad I_0\equiv H_0\equiv\frac{1}{1-\sop}L_1
\end{align*}
where
\begin{align}
	H_N&\equiv\sum_{p=0}^{N}T_p\frac{\sop^p-\sop}{1-\sop}V_{N-p}H_0,\label{eq:hn-def}
\end{align}
which is the core of the new representation.
We will find that the quantity $H_N$ is the new representation, a sum over just one $h_N$. Furthermore, through $V_N$ we can relate the original and permutation representation, then verify that the extra function between the two vanishes under reversal. The binomial aspect to the number of permutations is controlled by $c_N\lop$ in the definition of $I_N$. The $c_N$ provides the next $\coth$ in the product and $\lop=1-\sop$ provides twice the terms from the previous order.

The next step is to find the recurrence relation~\eqref{eq:v-rec} in terms of $H_N$ instead of $V_N$ and solve it for this quantity. We complete the rest of the analysis with generating functions and so we define
\begin{align*}
	H(z)\equiv\sum_{N=0}^{\infty}H_Nz^N,\quad I(z)\equiv\sum_{N=0}^{\infty}I_Nz^N.
\end{align*}
To turn equation~\eqref{eq:hn-def} into generating functions, multiply by $z^N$, sum over $N$ and reordering the sums on the right-hand side to achieve
\begin{align}
	H(z)&=\sum_{p=0}^{\infty}T_p\frac{\sop^pz^p-\sop z^p}{1-\sop}\sum_{N=p}^{\infty}V_{N-p}z^{N-p}H_0\nonumber\\
	&=\frac{T(\sop z)-\sop T(z)}{1-\sop}V(z)H_0\equiv W(z,\sop z)^{-1}V(z)H_0\label{eq:hv}
\end{align}
where we generalise the previous generating function $W(x)$ to a new central quantity $W(x,y)$ which we can write in a variety of ways,
\begin{align}
	W(x,y)^{-1}&=\frac{xT(y)-yT(x)}{x-y}=\frac{xy}{x-y}(\coth y-\coth x)\nonumber\\
	&=\frac{xy}{(x-y)}\frac{\sinh(x-y)}{\sinh x \sinh y}=\frac{G_1(x)G_1(y)}{G_1(x-y)}=W(y,x)^{-1}\label{eq:reciprocal-w},
\end{align}
and obviously has the $x\leftrightarrow y$ symmetry. The function $W(x,y)$ is almost identical to the generating function defined in Appendix E of the previous paper that was used to generate $f_n$ and to remove apparent singularities in $G_{N}$. This generating function is a polynomial in $\sop$ at each order in $z$ that is symmetric around a central point. This encodes the reversal symmetry and is represented mathematically by $z\leftrightarrow \sop z$ or, $W(z,\sop z)=W(\sop z,z)$, telling us that the structure of the coefficients is $W_{m,N-m}=W_{N-m,m}$ where $W_{m,N-m}$ is the coefficient of $z^m (\sop z)^{N-m}$ and so on in the Taylor series.
This includes the previous definitions as limiting cases
\begin{align*}
	W(x,x)\equiv W(x),\implies	W(z,z)=W(z),\quad W(\sop z,\sop z)= W(\sop z).
\end{align*}

Now we perform the final few algebraic manipulations to reach the permutation representation.
We already considered the right-hand side of~\eqref{eq:v-rec} applied to $H_0=(1-\sop)^{-1}L_1$, now consider the full equation: multiply by $z^N$ and sum over $N$ to find~\eqref{eq:v-rec} in terms of generating functions,
\begin{align}
	T(\sop z)T(z)V(z)H_0-\sop V(z)H_0z^2=I(z)\label{eq:v-rec-gen}
\end{align}
and we want to eliminate the symmetric partition representation $V(z)H_0$ in favour of $H(z)$, then solve the resulting equation to find that $H(z)$ is the new representation.
Now we simplify the hyperbolic prefactors on the left-hand side of~\eqref{eq:v-rec-gen}.
We can write the left-hand side as $(T(\sop z)T(z)-\sop z^2) V(z)H_0$ and use equation~\eqref{eq:hv} to eliminate $V(z)H_0$ in favour of $W(z,\sop z)H(z)$. The hyperbolic prefactor is then $(T(x)T(y)-xy)W(x,y)$ with $x=z$ and $y=\sop z$, which we can simplify using hyperbolic algebra into
\begin{align*}
	(T(x)T(y)-xy)W(x,y) &= xy\frac{(x-y)}{xy}\frac{\coth(y)\coth(x)-1}{\coth(y)-\coth(x)}=(x-y)\coth(x-y)=T(x-y),
\end{align*}
which for $x=z$, $y=\sop z$ we find $T((1-\sop)z)=T(\lop z)$. Multiplying through by the inverse, $[(T(z)T(\sop z)-\sop z^2)W(z,\sop z)]^{-1}$, isolates $H(z)$ on the left-hand side to provide the main result
\begin{subequations}
	\begin{align}
		H(z) &= t(\lop z)I(z)\\
		H_N &= \sum_{p=0}^{N-1}t_p\lop^p c_{N-p}\lop H_{N-1-p}+t_N\lop^N H_0,\quad \lop H_0\equiv L_1
	\end{align}
\end{subequations}
where in the second line we used the definition of the transient $I_N$ to eliminate it in favour of $H_{N-1}$, achieving a formula purely in terms of $H_N$, the first few are given in equations~\eqref{eq:H1} to~\eqref{eq:H4} in the introduction.
One can recognise this formula as a linear combination of reversal operators applied to one $h_r$ through the following argument. Consider the rightmost term, $\dots\lop\coth(L_1)$, in equations~\eqref{eq:H1} to~\eqref{eq:H4} and higher-order analogues, since all terms to its left depend on $L_1+L_2+\cdots$, we recognise that the raising operator becomes a cyclic permutation, $\lop\coth(L_1)=\coth(L_1)-\coth(L_2)=(1-R_2R_1)\coth(L_1)$ and commute $(1-R_2R_1)$ through the other terms as they are invariant under the cyclic permutation of two indices since they contain the sum of at least two indices, this amounts to $h(L_1+L_2+\cdots)\lop\coth(L_1)=(1-R_2R_1)h(L_1+L_2+\cdots)\coth(L_1)$. Continuing this without reordering the factors $(1-R_mR_{m-1})$ provides $P_N$ as the linear combination of permutations (equation~\eqref{eq:pn-intro}). What remains is the product of $\coth$'s with gaps given by $t_p$'s that one finds after repeatedly using edge identities to break down an $h_n$ into single $\coth$'s. Using the edge identities in reverse to build the $h_n$ again provides the alternative statement of the main result,
\begin{align}
	H_N = P_N\, h_N(L_1,L_1+L_2,\dots,L_1+L_2+\cdots+L_N)L_1
\end{align}
as what we refer to as the permutation representation. This representation involves a sum over only a single $h_N$, completing the progression in Section~\ref{sec:edge-K} from three $h_r$'s in a product to two, and now to just one.

We can now write the original representation in terms of the permutation representation. Eliminate $V(z)$ in~\eqref{eq:ev} in favour of $H(z)$ by using their relation given in~\eqref{eq:hv},
\begin{align}
	E(z) &= T(\sop z)W(z,\sop z)H(z)-s(z)H_0+\Delta(z) \label{eq:gen-ef}.
\end{align}
We have now related the original representation $E(z)$ to the new representation $H(z)$.
In order to show that $G_{N+1}$ is generated by $H_N$, we need to show that $X_N=E_N-H_N$ vanishes under reversal and we do this in the next section.

\section{Proof of equivalence}\label{sec:remainder}
In this section we prove that the permutation representation is equivalent to the original representation at the level of $G_N$ by proving that the additional function vanishes under the appropriate symmetric linear combination. We use the symmetric linear combination of the two partition representations given in equation~\eqref{eq:vkk}, which is invariant under the appropriate reversal, to prove equivalence. 
The proof takes place for each order individually because we rely on reversal properties of a function of a given number of variables.
An alternative proof of equivalence, using the permutation representation, is provided in Appendix~\ref{app:sec-equiv}.

We will prove that 
\begin{align}
	\sinh(L_1+L_2+\cdots+L_{N+1})&G_{N+1}(L_1,L_2,\dots,L_{N+1})=\lop H_N\nonumber\\
	&=(1-(-1)^{N+1}R_{N+1})H_{N}\label{eq:new-gn}
\end{align}
for $N\geq 1$, note that the $N=0$ case is catered for by definition. This requires proving
\begin{align}
	(1-(-1)^{N+1}R_{N+1})E_{N}=(1-(-1)^{N+1}R_{N+1})H_{N}
\end{align}
for $N\geq 1$. The zeroth order terms in both $E(z)$ and $H(z)$ produce the correct and identical $G_N$ by definition since we defined $2E_0\equiv L_1$ and $\lop H_0\equiv L_1$ and each one is defined in such a way to be internally consistent with the rest of their respective representation. Equation~\eqref{eq:gen-ef} relates $E_N$ and $H_N$ and we see that they are not strictly equivalent, but since only $G_{N+1}$ must be conserved, we just require that the additional function, say $X_{N}$, has the symmetry required to cancel out upon taking the sum/difference. To this end we define the extra function by
\begin{align}
	E_{N}\equiv H_{N}+X_{N},
\end{align}
where the extra function can in principle depend on all present variables but we find that it only depends on the variables shared with the other half that forms $G_{N+1}$. We shall now prove that
\begin{align}
	(1-(-1)^{N+1}R_{N+1})X_N=0\label{eq:xn-van},
\end{align}
so that $H_N$ is a representation of $G_{N+1}$.

The extra function is
\begin{align}
	X(z) &= E(z)-H(z)=\Delta(z)-s(z)H_0+[T(\sop z)-W(z,\sop z)^{-1}]V(z)H_0\label{eq:x}
\end{align}
where we used $H(z)=W(z,\sop z)^{-1}V(z)H_0$ in the final equality and we found that substituting~\eqref{eq:vkk} in place of $V(z)H_0$ leads to a more straightforward proof.

The hyperbolic part of the original $V(z)H_0$ is multiplied by
\begin{align}\label{eq:three-hyp}
	\frac{\sop}{1-\sop}(T(z)-T(\sop z))=\sop(T(z)-W(z,\sop z)^{-1})=T(\sop z)-W(z,\sop z)^{-1},
\end{align}
where we used the reciprocal of $W(z,\sop z)$ given in equation~\eqref{eq:reciprocal-w} to carry out the hyperbolic algebra and we use the final expression later in the proof.
We use the three hyperbolic representations of this quantity when considering the three pieces of~\eqref{eq:vkk}, $(J(z)+K(z))/2$, $W(z)H_0/2$ and $W(\sop z)H_0/2$ respectively.
Substituting equation~\eqref{eq:vkk} into~\eqref{eq:x} and using the three hyperbolic representations of~\eqref{eq:three-hyp} we find that~\eqref{eq:x} can be written as
\begin{equation}\label{eq:x-inter}
	\begin{aligned}
		X(z)&=\frac{\sop}{1-\sop}(T(z)-T(\sop z))\frac{1}{2}(J(z)+K(z))+\Delta(z)-s(z)H_0\\
		&\quad +\frac{1}{2}\sop(T(z)-W(z,\sop z)^{-1})W(z)H_0+\frac{1}{2}(T(\sop z)-W(z,\sop z)^{-1})W(\sop z)H_0.
	\end{aligned}
\end{equation}
Next we use that
\begin{align*}
	\Delta(z)=\frac{1}{2}s(z)L_1+\frac{1}{2}\frac{s(z)-s(\sop z)}{1-\sop}L_1=\frac{1}{2}\left(2s(z)-\sop s(z)-s(\sop z)\right)H_0
\end{align*}
to cancel three of the linear terms in equation~\eqref{eq:x-inter}, leaving
\begin{equation}\label{eq:x-starting}
	\begin{aligned}
		X(z)&=\frac{\sop}{1-\sop}(T(z)-T(\sop z))\frac{1}{2}(J(z)+K(z))\\
		&\quad -W(z,\sop z)^{-1}\frac{1}{2}(\sop W(z)+W(\sop z))H_0
	\end{aligned}
\end{equation}
as the form of the extra function that we use for the proof. We now deal with the hyperbolic terms, then with the linear terms.

The hyperbolic terms are controlled by the operator
\begin{align*}
	\frac{1}{2}\frac{\sop}{1-\sop}(T(z)-T(\sop z))=\frac{1}{2}\sum_{p=2}^{\infty}T_pz^p\frac{\sop-\sop^{p+1}}{1-\sop}=\frac{1}{2}\sum_{p=2}^{\infty}T_pz^p\sum_{r=1}^{p}\sop^r
\end{align*}
which is applied to $J(z)+K(z)$ to give
\begin{align}
	\frac{1}{2}\sum_{p=2}^{\infty}T_pz^p\sum_{r=1}^{p}\sop^r\sum_{m=1}^{\infty}z^m(J_m+K_m)=
	\frac{1}{2}\sum_{N=3}^{\infty}z^N\sum_{p=0}^{N-1}T_p\sum_{r=1}^{p}\sop^r(J_{N-p}+K_{N-p})\label{eq:xh}
\end{align}
with $m+p=N$.
We now must apply $(-1)^{N+1}R_{N+1}$ to this expression at order $z^N$, which is given by $\sum_{p=0}^{N-1}T_p\sum_{r=1}^{p}\sop^r(J_{N-p}+K_{N-p}).$ Consider 
\begin{align*}
	\sop^rJ_m(L_1,\dots,L_m)=J_m(L_{1+r},\dots,L_{m+r})
\end{align*}
for $N+2>m+r$, then
\begin{align*}
	(-1)^{N+1}R_{N+1}J_m(L_{1+r},\dots,L_{m+r})&=(-1)^{N+1}J_m(L_{N-r+1},\dots,L_{N+2-m-r})\\
	&=(-1)^{N+1}\sop^{N-m-r+1}J_m(L_{m},\dots,L_{1})
\end{align*}
leading to the result
\begin{align}
	(-1)^{N+1}R_{N+1}\sop^rJ_m= (-1)^{N+1}\sop^{N-m-r+1}\tilde J_m\label{eq:rj}
\end{align}
and similarly for $K_m$. From equation~\eqref{eq:xh} we have $m=N-p$
\begin{align*}
	(-1)^{N+1}R_{N+1}\sop^r (J_{N-p}+K_{N-p})=(-1)^{N+1}\sop^{p-r+1}(\tilde J_{N-p}+\tilde K_{N-p})
\end{align*}
Now we use this result and consider our quantity with reversed arguments,
\begin{align*}
	\tilde J_N+\tilde K_N &=(-1)\sum_{m=0}^{N}(-1)^m h_m(L_{N+1-m},\dots,L_{N+1-m}+\cdots+L_N)h_{N-m}(L_{N-m},\dots,L_1+\cdots+L_{N-m})\nonumber\\
	&\quad\times(L_N+\cdots+L_{N+1-m}-[L_{N-m}+\cdots+L_1])
\end{align*}
and then set $n+m=N$ to find
\begin{align*}
	&\sum_{n=0}^{N}(-1)^{N-n}h_n(L_n,\dots,L_1+\cdots+L_n)h_{N-n}(L_{1+n},\dots,L_{1+n}+\cdots+L_N)\nonumber\\
	&\quad\times(L_1+\dots+L_n-[L_{1+n}+\cdots+L_N])=(-1)^{N+1}(J_N+K_N)
\end{align*}
and since $\sum_{p=1}^{r}\sop^{p-r+1}=\sum_{s=1}^{p}\sop^s$, which one can see using $s+r=p+1$, the result is proven for the hyperbolic terms.

We now consider the linear terms---the second line of~\eqref{eq:x-starting}---which are given by
\begin{align}
	X^l(z,\sop z)L_1&\equiv-\frac{G_1(z)G_1(\sop z)}{G_1(z-\sop z)}\frac{1}{2}\left[\frac{\sop}{G_1(z)^2}+\frac{1}{G_1(\sop z)^2}\right]\frac{1}{1-\sop}L_1\nonumber\\
	&=-\frac{1}{G_1(z-\sop z)}\frac{1}{2}\left[\frac{z\sop G_1(\sop z)}{G_1(z)}+\frac{zG_1(z)}{G_1(\sop z)}\right]\frac{1}{z-\sop z}L_1\nonumber\\
	&\equiv X_0^l+\sum_{N=1}^{\infty}X^l_Nz^NL_1\nonumber\\
	X^l_N&\equiv \sum_{r=1}^{N-1} X^l_{r,N-r}\sop^{N-r},\quad N>0\label{eq:xl-sum}
\end{align}
where we used~\eqref{eq:reciprocal-w} to write $W^{-1}(z,\sop z)$ in terms of $G_1$'s, and clearly the function is antisymmetric under swapping its arguments, $X^l(z,\sop z)+X^l(\sop z,z)=0$. This antisymmetry is crucial and it specifies the symmetry of the coefficients in this generating function as
\begin{align*}
	X^l_{r,N-r}=-X^l_{N-r,r},
\end{align*}
for $N\geq 1$ and $1\leq r\leq N-1$, which we use shortly. 
The zeroth order term is singular because the new representation is at the level of $H_N$ due to the presence of $H_0$, explicitly the zeroth order term is $X^l_{0,0}=1/2-1/(1-\sop)$, and we separate this from the analysis. This fact is irrelevant to the current discussion because we only require to produce the correct quantity at the level of $G_N$.

Next we consider the linear terms at order $N$ and apply $(-1)^{N+1}R_{N+1}$. We find that the linear terms are invariant under this operation by using the antisymmetry of the generating.
First we can evaluate the alternating minus sign as $(-1)^{N+1}\mapsto -1$ since only even $N$ contribute to the linear terms piece.
Next we determine the resulting action of a shift followed by the reversal,
\begin{align*}
	R_{N+1}\sop^{N-r}L_1=R_{N+1}L_{N-r+1}=L_{r+1}=\sop^r L_1,
\end{align*}
and with this we can proceed to prove the equivalence. Consider $(-1)^{N-1}R_{N+1}$ applied to the linear terms at order $N$,
\begin{align*}
	(-1)^{N+1}R_{N+1}X^l_NL_1=	-R_{N+1}\sum_{r=1}^{N-1}X^l_{r,N-r}\sop^{N-r} L_1
\end{align*}
where in the final expression we evaluated the alternating minus sign and substituted the form of $X^l_N$ given in~\eqref{eq:xl-sum}. Next we use the result of a shift then reversal so that the previous expression becomes
\begin{align*}
	-\sum_{r=1}^{N-1}X^l_{r,N-r}\sop^{r} L_1.
\end{align*}
The penultimate step is to use the antisymmetry property $X^l_{r,N-r}=-X_{N-r,r}$
to find
\begin{align*}
	=-\sum_{r=1}^{N-1}(-X^l_{N-r,r})\sop^{r} L_1
\end{align*}
and finally to use $r'=N-r$ in the sum
\begin{align}
	\sum_{r'=1}^NX^l_{r',N-r'}\sop^{N-r'}L_1 =X^l_NL_1\label{eq:xl-res}
\end{align}
where we recognised $X^l_N$ from its definition in~\eqref{eq:xl-sum}, and the reversal operator acts to swap the arguments of $X^l(z,\sop z)$.
This completes the proof of equation~\eqref{eq:xn-van} and therefore $H_N$ is a representation of $G_{N+1}$ in the sense of equation~\eqref{eq:new-gn}.

In the next section we prove that $G_{N+1}$ obey marching identities by proving that $H_N$ do and then that $G_{N+1}$ inherit this property.

\section{Proof of marching identities}\label{sec:marching-identities}
This section provides a proof of marching identities, our strategy is to prove that $H_{N}$ obey marching identities then trivially extend this to $G_{N+1}$.
Applying a marching identity is nothing more than applying a sum of permutation operations to $H_{N}$ and checking the resulting combination of $h_{N}$ cancels out. We do this at the operator level and show that it is the marching permutations combined with the permutation structure of $H_{N}$ that cancels out rather than anything hyperbolic! Therefore this also proves that the denominator analogue in Appendix~\ref{app:denom} has marching identities.

The marching identities can be specified with a sum of permutation operators $M_{N,m}$ which describes marching $m$ out of $N$ parameters through the string. To fourth order these operators have the action
\begin{equation*}
	\begin{aligned}
		M_{1,1}[L_1] &= [L_1]\\
		M_{2,1}[L_1L_2]&=[L_1L_2]+[L_2L_1]\\
		M_{3,1}[L_1L_2L_3]&=[L_1L_2L_3]+[L_2L_1L_3]+[L_2L_3L_1]\\
		M_{4,1}[L_1L_2L_3L_4]&=[L_1L_2L_3L_4]+[L_2L_1L_3L_4]+[L_2L_3L_1L_4]+[L_2L_3L_4L_1]\\
		M_{3,2}[L_1L_2L_3]&=[L_1L_2L_3]+[L_1L_3L_2]+[L_3L_1L_2]\\
		M_{4,2}[L_1L_2L_3L_4]&=[L_1L_2L_3L_4]+[L_1L_3L_2L_4]+[L_1L_3L_4L_2]\\
		&\quad+[L_3L_1L_2L_4]+[L_3L_1L_4L_2]+[L_3L_4L_1L_2]\\
		M_{4,3}[L_1L_2L_3L_4]&=[L_1L_2L_3L_4]+[L_1L_2L_4L_3]+[L_1L_4L_2L_3]+[L_4L_1L_2L_3].
	\end{aligned}
\end{equation*}
Proving that the $H_{N}$ obey marching identities amounts to proving $M_{N,m}P_N=0$. We prove this by induction first for $m=1$ then for all $m$.

For this section of the proof we require the recursive definition of $M_{N,1}$ and $M_{N,N-1}$
\begin{align}
	M_{N,1} &= M_{N-1,1}+(12\cdots N-1N)\\
	M_{N,N-1} &= 1+(N-1 N)M_{N-1,N-2}
\end{align}
where $M_{N,1}$ can be thought of marching the first parameter through $N-1$ parameters with the previous marching operator, then including the final permutation involves putting the first parameter at the end and shifting the others back by one. We think of $M_{N,N-1}$ in an analogous way where $1$ provides the identity permutation and the other permutations are provided by marching the first $N-2$ parameters through the first $N-1$ of them and then including the fact that we have the new, final variable with $(N-1 N)$. Equivalently we can think of moving $L_{N-1}$ sequentially past the others and then swapping $L_N$ with $L_{N-1}$ to make it that $L_N$ moved past the others.
We also require the definitions $M_{N,0}\equiv 1$, $M_{N,N}\equiv 1$ and the facts
\begin{align}
	R_N M_{N,m}R_N&=M_{N,N-m}\\
	(NN-1\cdots 21)&=R_{N-1}R_N, \text{ and the inverse }(12\cdots N-1N)=R_NR_{N-1}, \label{eq:fact3}
\end{align}
where we have used the reversal operators from the previous section. The first fact is the analogue of a similarity transform in linear algebra. The second fact can be understood as reversing all parameters, putting the final one at the front, then undoing the reversal on the other $N-1$ parameters putting them back in their original order, having the effect of moving the final parameters to the start. The inverse has the effect of moving the first parameter to the end by analogous logic. Examples make this obvious
\begin{align*}
	R_3R_4[L_1L_2L_3L_4] &= R_3[L_4L_3L_2L_1]=[L_4L_1L_2L_3]\\
	R_4R_3[L_1L_2L_3L_4] &= R_4[L_3L_2L_1L_4]=[L_2L_3L_4L_1].
\end{align*}

We now use mathematical induction to prove that $H_N$ obey marching identities. We first show $M_{2,1}P_2=0$. This is trivially $$M_{2,1}P_2=(1+R_2)(1-R_2)=1-R_2+R_2-(R_2)^2=0,$$ and so inductively if $M_{N-1,1}P_{N-1}=0$ and $M_{N-1,N-2}P_{N-1}=0$ then using equation~\eqref{eq:pn-intro} provides 
\begin{subequations}
	\begin{align}
		M_{N,1}P_N&=M_{N,1}P_{N-1}-(-1)^NR_N(R_NM_{N,1}R_N)P_{N-1}\\
		&= M_{N,1}P_{N-1}-(-1)^NR_NM_{N,N-1}P_{N-1}\\
		&= M_{N-1,1}P_{N-1}+ (12\cdots N)P_{N-1}\nonumber\\
		&\quad -(-1)^NR_NP_{N-1}-(-1)^NR_N(N-1N)M_{N-1,N-2}P_{N-1}\\
		&=(12\cdots N)\left[1+(-1)^{N-1}R_{N-1}\right]P_{N-1}\label{eq:m1}
	\end{align}
\end{subequations}
where in the first equality we inserted $R_N^2=1$, in the second equality we performed the similarity transform on $M_{N,1}$, in the third equality we used the recursive definitions of $M_{N,m}$ and the inductive assumptions, in the final equality we used equation~\eqref{eq:fact3}; and the right-hand side is zero by the symmetry of $P_{N-1}$. We also have the analogous
\begin{subequations}
	\begin{align}
		M_{N,N-1}P_N&=M_{N,N-1}P_{N-1}-(-1)^NR_N(R_NM_{N,N-1}R_N)P_{N-1}\\
		&= M_{N,N-1}P_{N-1}-(-1)^NR_NM_{N,1}P_{N-1}\\
		&= P_{N-1}+ (N-1\; N)M_{N-1,N-2}P_{N-1}\nonumber\\
		&\quad -(-1)^NR_N(M_{N-1,1}+(12\cdots N))P_{N-1}\\
		&=P_{N-1}-(-1)^NR_NR_NR_{N-1}P_{N-1}=\left[1+(-1)^{N-1}R_{N-1}\right]P_{N-1}.
	\end{align}
\end{subequations}

Now for all $m$, including the two previous cases, we use the relationship 
\begin{align}
	M_{N,m}=M_{N-1,m}+(mm+1m+2\cdots N)M_{N-1,m-1}
\end{align}
which we interpret as marching $m$ parameters through $N-1-m$ of them and $L_N$ remains at the end, plus the contribution when $L_m$ has passed $L_N$ and becomes the final variable; which we write as a lower order identity with a permutation to put $L_m$ at the end. The inductive step is
\begin{subequations}
	\begin{align}
		M_{N,m}P_{N} &= M_{N,m}P_N-(-1)^NR_NM_{N,N-m}P_{N-1}\\
		&=M_{N-1,m}P_{N-1}+(mm+1m+2\cdots N)M_{N-1,m-1}P_{N-1}\nonumber\\
		&\quad-(-1)^NR_N\left[(N-mN-m+1\cdots N)M_{N-1,N-m-1}+M_{N-1,N-m}\right]P_{N-1}
	\end{align}
\end{subequations}
which vanishes by the induction assumption which provide lower order marching identities; except the special cases of $m=1$ where
\begin{align}
	M_{N,1}P_N &= M_{N-1,1}P_{N-1}+(12\cdots N)M_{N-1,0}P_{N-1}\nonumber\\
	&\quad -(-1)^NR_N\left[(N-1N)M_{N-1,N-2}+M_{N-1,N-1}\right]P_{N-1}\nonumber\\
	&=\left[(12\cdots N)-(-1)^NR_N\right]P_{N-1}
\end{align}
which is zero in the same manner of equation~\eqref{eq:m1}. The special case of $m=N-1$ is 
\begin{align}
	M_{N,N-1}P_N &= M_{N-1,N-1}P_{N-1}+(N-1N)M_{N-1,N-2}P_{N-1}\nonumber\\
	&\quad -(-1)^NR_N\left[(12\cdots N)M_{N-1,0}+M_{N-1,1}\right]P_{N-1}\nonumber\\
	&=\left[1-(-1)^NR_N(12\cdots N)\right]P_{N-1}
\end{align} which is also zero in the same manner of equation~\eqref{eq:m1}. 
This completes the proof that $H_N$ obey marching identities.

Extending this to $G_{N+1}$ requires writing the main result as
\begin{align}
	\sinh(L_1+L_2+\cdots+L_{N+1})G_{N+1}(L_1,L_2,\dots,L_{N+1})&= (1-(-1)^{N+1}R_{N+1})H_N\nonumber\\
	&=(1-(-1)^{N+1}R_{N+1})P_{N}\, h_{N}L_1\nonumber\\
	&=P_{N+1}\,h_{N}L_1 \label{eq:gn-pn}
\end{align}
where the right-hand side of the first line is valid for $N>0$ and the final line includes the $N=0$ case. Marching identities amount to applying $M_{N+1,m}$ to equation~\eqref{eq:gn-pn} and we have already proved $M_{N+1,m}P_{N+1}=0$, so $G_{N+1}$ obeys marching identities.

\section{Conclusion}
A permutation based power series representation of the Baker-Campbell-Hausdorff formula has been found by using generating functions and the $\coth$ angle addition identity.
The permutation representation has a structure that permits proof of marching identities from the previous paper \cite{moodie2021} and these are crucial to doing perturbation theory with operators related by the BCH formula.
The intermediate representations that are controlled by $V(z)$, $J(z)$ and $K(z)$ all have properties of their own which may prove useful in proving other properties of the series or its coefficients.
Introducing the critical operators $\sop$ and $\lop=1-\sop$ allowed us to make each term in the resulting sum start from the first or last variable at a given order; making the structure of each term identical and therefore susceptible to generating functions.

The main mathematical purpose of this paper is to replace the original representation for $G_{N+1}$ by the new, elementary linear sum of the central $h_N$ quantities, which we believe provides a better starting point for applications. Indeed, we used this representation to prove marching identities.

\subsubsection*{Acknowledgements}
We thank Sam Pickering and Thomas Edge for comments on the manuscript.

\appendix
\section{Future use of the formula}\label{app:pert}
In this appendix we present the main result explicitly to third order. Then we consider a diagonal $A$ operator and non-diagonal $B$ which provides simpler expressions and a ready-to-use formula. Finally we present the perturbation theory that one can derive using the power series representation.

We first restate the main result~\eqref{eq:resPH} with the terms written explicitly to third order,
\begin{align}
	C &= A + \frac{L_1}{\sinh(L_1)}B_1 + \frac{1}{\sinh(L_1+L_2)}\big[L_1h_1(L_1)-L_2h_1(L_2)\big]B_1 B_2\nonumber\\
	&\quad+\frac{1}{\sinh(L_1+L_2+L_3)}\bigg[L_1h_2(L_1,L_1+L_2)-L_2h_2(L_2,L_1+L_2)\nonumber\\
	&\qquad\qquad\qquad\qquad\qquad -L_2h_2(L_2,L_2+L_3)+L_3h_2(L_3,L_2+L_3)\bigg]B_1B_2B_3+\cdots\\
	h_1(x)&=\coth(x)\nonumber\\
	h_2(x,y)&= \coth(x)\coth(y)-\frac{1}{3}\nonumber
\end{align}
where we provide the relevant $h_r$ to make this appendix self-contained, and we can consider working in a basis where $A$ is diagonal, representing an exactly solved Hamiltonian, and the other quantity, $B$, is not diagonal in this basis. Then we recall that a Taylor-expandable function, say $g$, of the $L_m$ operators can be written as
\begin{align*}
	\mel*{n}{g(L)B}{n'}=g(a_n-a_{n'})B_{nn'},\quad \mel*{n}{g(L_1,L_2)B_1B_2}{n'}=\sum_{n_1}g(a_n-a_{n_1},a_{n_1}-a_{n'})B_{nn_1}B_{n_1n'}
\end{align*}
etc., where $A_{nn'}\equiv a_n\delta_{nn'}$ defines notation for the eigenvalues $a_n$ of the diagonal operator $A$, $B_{nn'}\equiv\mel*{n}{B}{n'}$, and the second example requires inserting a complete basis of $A$ between the product of $B$'s.
Then to second order the matrix elements of $C$ are
\begin{align}
	&\mel*{n}{C}{n'}=a_n\delta_{nn'}+\frac{(a_n-a_{n'})}{\sinh(a_n-a_{n'})}B_{nn'}\nonumber\\
	&\quad + \frac{1}{\sinh(a_n-a_{n'})}\sum_{n_1}\left[(a_n-a_{n_1})h_1(a_n-a_{n_1})-(a_{n_1}-a_{n'})h_1(a_{n_1}-a_{n'})\right]B_{nn_1} B_{n_1n'}\nonumber\\
	&\quad+\frac{1}{\sinh(a_n-a_{n'})}\sum_{n_1,n_2}\bigg[(a_n-a_{n_1})h_2(a_n-a_{n_1},a_n-a_{n_2})\nonumber\\
	&\qquad-(a_{n_1}-a_{n_2})h_2(a_{n_1}-a_{n_2},a_n-a_{n_2})-(a_{n_1}-a_{n_2})h_2(a_{n_1}-a_{n_2},a_{n_1}-a_{n'})\nonumber\\
	&\qquad+(a_{n_2}-a_{n'})h_2(a_{n_2}-a_{n'},a_{n_1}-a_{n'})\bigg]B_{nn_1} B_{n_1n_2}B_{n_2n'}+\cdots,
\end{align}
where the arguments of all hyperbolic functions are now just numbers and the sum over $n_1$ is over a complete set of eigenstates of $A$. This equation is directly applicable to physical problems, and is indeed regular when one targets the diagonal components, $n=n'$. The differences in eigenvalues of the diagonal operator, $a_n-a_{n_1}$, can be thought of as analogues of the denominators in perturbation theory.

Eigenvalues and their gaps are central to physics therefore we are motivated to find the eigenvalues of the resulting operator $C$ from the BCH formula using the new power series representation. This was also mentioned in Appendix~A of the previous paper and will be the subject of a future paper. Indeed, the results in this paper are all necessary prerequisites to deriving perturbation theory for the operators that are related by the BCH formula. 

To derive perturbation theory using the power series one requires marching identities and the overcomplete representation of $G_N$, provided in Appendix~\ref{app:overcomplete}.
One can apply formal perturbation theory to the power series, for a diagonal $A$ and perturbative $B$. Interpreting each power of $B$ as a perturbing Hamiltonian, one can find the eigenvalues of $C$,
\begin{equation}
\begin{aligned}
	\mel*{n}{C}{n}\equiv c_n &\equiv c_n^{(0)} + c_n^{(1)} + c_n^{(2)} + c_n^{(3)} + c_n^{(4)} + \cdots\\
	c_n^{(0)} &= a_n\\
	c_n^{(1)} &= B_{nn}\\
	c_n^{(2)} &= \sum_{n_1}h_1(a_n-a_{n_1})B_{nn_1}B_{n_1n}\\
	c_n^{(3)} &= \sum_{n_1,n_2}h_2(a_n-a_{n_1},a_n-a_{n_2})B_{nn_1}B_{n_1n_2}B_{n_2n}\\
	&\quad-\sum_{n_1}h_2(a_n-a_{n_1},a_n-a_{n_1})B_{nn_1}B_{n_1n}B_{nn}
\end{aligned}
\end{equation}
where $a_n$ are again the eigenvalues of the diagonal operator, $n_m\neq n$, and this resulting formula is \textit{exactly} like ordinary perturbation theory when one has $C=A+B$, but the energy denominators have mapped to our $h_n$ functions which are provided above and in the main text in equations~\eqref{eq:h0} to~\eqref{eq:h6}.
\section{Algebra of number-convolution expressions}\label{app:number-gen}
Some of the relations in the main text contain $\sop^{-m}$ at the right-hand side which makes the expression not separable into generating functions. We call these number-convolutions as explained below equation~\eqref{eq:reversed-edge}. In this appendix we show how to invert these expressions in a way analogous to the method of recognising a convolution in generating function algebra.
We show how to do this with $T(z)=z\coth(z)$ and $t(z)=\tanh(z)/z=1/T(z)$ which generate the coefficients $T_p$ and $t_p$ respectively. These are inverses and this amounts to
\begin{align*}
	\sum_{p=0}^{N}T_pt_{N-p}=\delta_{N,0}=\sum_{p=0}^Nt_pT_{N-p}.
\end{align*}
The \textit{number} analogues are also inverses and are defined as operating on some quantity $X_N$ as
\begin{align*}
	\sum_{p=0}^NT_p\sop^p X_{N-p}\sop^{-p},\quad\sum_{p=0}^Nt_p\sop^p X_{N-p}\sop^{-p}
\end{align*}
which is only an operator if $X_N$ is since $\sop$ acts to only change the values of $L_m$ in the arguments.
Applying both number-convolutions to a quantity provides
\begin{align*}
	\sum_{p=0}^NT_p\sop^p \sum_{q=0}^{N-p}t_q\sop^q X_{N-p-q}\sop^{-q}\sop^{-p},
\end{align*}
and this provides
\begin{align*}
	\sum_{p=0}^{N}T_p\sum_{r=p}^{N}t_{r-p}\sop^rX_{N-r}\sop^{-r}=\sum_{r=0}^{N}\left(\sum_{p=0}^{r}T_pt_{r-p}\right)\sop^rX_{N-r}\sop^{-r}=\sum_{r=0}^{N}\delta_{r,0}\sop^r X_{N-r}\sop^{-r}=X_N
\end{align*}
where we used $p+q=r$ and that $t_{r-p}$ commutes with $\sop$ to achieve the first expression, then swapped the order of summation, factorised the sum over $p$ and evaluated it, to verify that these two operations are indeed inverses.
\section{Alternative proof of equivalence}\label{app:sec-equiv}
In this appendix we prove the equivalence of representations using the permutation representation, in contrast to the main text where we elected to use $V(z)H_0$ in the additional function. This is only slightly different to the procedure in the main text.
The proof is completed at the level of permutation operators, using $H_N$ as a basis, and the underlying hyperbolic details of this basis are irrelevant. One only requires the number of arguments the permutation operators are acting on so that we can find the commutation of $R_n$ and $\sop^m$, and that $R_NH_N=(-1)^{N+1}H_N$.

First define the relevant quantity
\begin{align*}
	X(z)\equiv E(z)-H(z)\equiv (T(\sop z)W(z,\sop z)-1)H(z)-s(z)H_0+\Delta(z)
\end{align*}
and to isolate a hyperbolic and linear piece we write $F(z)\equiv H(z)-H_0$ and associate the $H_0$ with the linear piece, giving
\begin{align*}
	X(z)&\equiv X^h(z,\sop z)F(z)+X^l(z,\sop z)L_1+X^l_0\\
	X^h(z,\sop z)&\equiv T(\sop z)W(z,\sop z)-1\equiv\sum_{N=1}^{\infty}X^h_Nz^N,\quad X^h_N\equiv \sum_{m=1}^{N-1} X^h_{N,N-m}\sop^{N-m}\\
	X^l(z,\sop z)&\equiv-s(z)\frac{1}{1-\sop}+\Delta(z)+T(\sop z) W(z,\sop z)\frac{1}{1-\sop} - \frac{1}{2}
\end{align*}
where $X^l_0=1/2-1/(1-\sop)$ is the zeroth order linear piece and contains the singularity, as in the main text, and $X^l(z,\sop z)$ is regular and its expansion starts at order $z^4$.
We complete the proof for the hyperbolic part of $X(z)$, the proof for the linear piece follows identical steps to that in the main text that arrives at equation~\eqref{eq:xl-res} so we do not repeat it.

The reversal structure is encoded in these generating functions as
\begin{align*}
	zX^h(z,\sop z)&=\sop z X^h(\sop z,z)\\
	X^l(z,\sop z)&+X^l(\sop z,z)=0
\end{align*}
which is the statement that the coefficients of the generating functions have the symmetry $X^h_{n+1,m}=X^h_{m,n+1}$ and antisymmetry $X^l_{n,m}=-X^l_{m,n}$ respectively. We also use that both $X^h(z,\sop z)$ and $X^l(z, \sop z)$ are each even in $z$. We are now ready to prove equation~\eqref{eq:xn-van}.

Consider the hyperbolic piece
\begin{align*}
	X^h(z,\sop z)F(z)&=\sum_{m=1}^{\infty}z^m\sum_{r=0}^{m-1}X^h_{r,m-r}\sop^{m-r} \sum_{q=0}^{\infty}F_{q}z^q,\quad F_0= 0\\
	&=\sum_{N=1}^{\infty}z^N\sum_{m=1}^{N}\sum_{r=0}^{m-1}X^h_{r,m-r}\sop^{m-r}F_{N-m}
\end{align*}
where we reordered the sums and used $q=N-m$.
We require the analogue of equation~\eqref{eq:rj}, to that end consider
\begin{align*}
R_{N+1}\sop^pF_n(L_1,\dots,L_n)=F_n(L_{N+1-p},\dots,L_{N+2-p-n})=\sop^{N+1-p-n}\tilde F_n=(-1)^{n+1}\sop^{N+1-p-n} F_n
\end{align*}
for $N+2>p+n$, where we used $\tilde F_n=F_n(L_n,\dots,L_1)=(-1)^{n+1}F_n(L_1,\dots,L_n)$ to reach the final equality; and we have $p=m-r$ and $n=N-m$, giving
\begin{align*}
	R_{N+1}\sop^{m-r}F_{N-m}=(-1)^{-m}\sop^{r+1}F_{N-m}=\sop^{r+1}F_{N-m}
\end{align*}
because $X^h(z,\sop z)$ is even in $z$ we used $(-1)^{-m}=1$. Applying $(-1)^{N+1}R_{N+1}$ to the hyperbolic piece at order $z^N$ then provides
\begin{align*}
	(-1)^{N+1}R_{N+1}\sum_{m=1}^{N}\sum_{r=0}^{m-1}X^h_{r,m-r}\sop^{m-r}F_{N-m}&=\sum_{m=1}^{N}\sum_{r=0}^{m-1}X^h_{r,m-r}\sop^{r+1}F_{N-m}\\
	&=\sum_{m=1}^{N}\sum_{r=0}^{m-1}X^h_{m-(r+1),r+1}\sop^{r+1}F_{N-m}\\
	&=\sum_{m=1}^{N}\sum_{r=0}^{m-1}X^h_{r,m-r}\sop^{m-r}F_{N-m}
\end{align*}
where in the final line we used the symmetry $X^h_{r,m-r}=X^h_{m-(r+1),r+1}$ and the final equality is equivalent to the starting point because the sum over $r$ is  identical to the initial one,
\begin{align*}
	\sum_{r=0}^{m-1}X^h_{m-(r+1),r+1}\sop^{r+1}=\sum_{r=0}^{m-1}X^h_{r,m-r}\sop^{m-r},
\end{align*}
and the result is proven for the hyperbolic piece.

Finally we give some examples of the extra function $X_N$ in terms of the permutation representation by expanding equation~\eqref{eq:gen-ef}, 
\begin{subequations}
\begin{align*}
	X_1 &= 0 \\
	X_2 &= \Delta_2 -\frac{1}{3}(2+\sop)L_1=0 \\
	X_3 &= \frac{1}{3}(\sop+\sop^2)F_1\\
	X_4 &= \frac{1}{3}(\sop+\sop^2)F_2+\frac{1}{45}(-6-7\sop-3\sop^2+\sop^3)L_1+\Delta_4\nonumber\\
	&=\frac{1}{3}(\sop+\sop^2)F_2-\frac{4}{45}(\sop-\sop^3)L_1\\
	X_5 &= \frac{1}{3}(\sop+\sop^2)F_3 -\frac{1}{45}(\sop-4\sop^2-4\sop^3+\sop^4)F_1\\
	X_6 &= \frac{1}{3}(\sop+\sop^2)F_4 -\frac{1}{45}(\sop-4\sop^2-4\sop^3+\sop^4)F_2\nonumber\\
	&\quad +\Delta_6L_1+(-\frac{2 \sop ^5}{945}+\frac{2 \sop ^4}{189}-\frac{2 \sop ^3}{315}-\frac{22 \sop ^2}{945}-\frac{2
		\sop }{189}-\frac{4}{315})\nonumber\\
	&= \frac{1}{3}(\sop+\sop^2)F_4 -\frac{1}{45}(\sop-4\sop^2-4\sop^3+\sop^4)F_2+\frac{4}{945} \sop  \left(\sop ^4+4 \sop ^3-4 \sop -1\right)L_1.
\end{align*}
\end{subequations}
\section{Denominator analogue}\label{app:denom}
This appendix gives a concrete physical understanding of how the hyperbolic algebra in the main text is combining the partition representation into the permutation representation.
We introduce a denominator analogue of a symmetric partition representation, $J(z)$, and prove that using a denominator identity and resumming this provides a linear combination of permutations generated by a sequence of reversal permutations applied to a single term that is analogous to $L_1 h_N$.

We next construct a denominator analogue of $J(z)$ which eliminates hyperbolic complications. We then resum the denominators into an analogue of the permutation representation. We do this by using a sequence of identities to combine denominators that are separated by the partition.
Mathematically one may think of this section as proving the leading order term in the permutation representation since the first term in the Taylor series of $\coth(z)$ is $1/z$.

The denominator analogue amounts to setting 
\begin{align*}
	h_n(x_1,x_1+x_2,\dots,x_1+x_2+\cdots+x_n)\mapsto \frac{1}{x_1 (x_1+x_2)\cdots(x_1+x_2+\cdots+x_n)}
\end{align*}
and so on, such that products become
\begin{align*}
	h_2(x_2,x_1+x_2)h_2(x_3,x_3+x_4)\mapsto\frac{1}{x_2}\frac{1}{x_1+x_2}\frac{1}{x_3+x_4}\frac{1}{x_3}
\end{align*}
and we write the order of the denominators in reverse for the second term, to have the two extremal factors next to each other. Of course, we could write $h_n$ in that way too if we wish since it has the symmetry $h_n(x_1,\dots,x_n)=h_n(x_n,\dots,x_1)$.

The fundamental denominator identity that we use is
\begin{align*}
	\frac{1}{a}\frac{1}{b}=\left(\frac{1}{a}+\frac{1}{b}\right)\frac{1}{a+b}
\end{align*} 
to write denominators which are separated by the partition, as denominators containing the sum of the arguments. The hyperbolic analogue is
\begin{align*}
	h_1(a)h_1(b)+1=(h_1(a)+h_1(b))h_1(a+b)
\end{align*}
which is nothing more than the angle addition formula for $\coth$.

We now prove that we can rewrite the denominator analogue as a linear combination of permutation operators applied to just one central quantity. Consider the symmetric partition formula $J_N$ provided in~\eqref{eq:k}, using $L_r\mapsto x_r$ as a basis, and its denominator analogue
\begin{align*}
	J_N &= \sum_{r=1}^{N}(-1)^{r-1}x_m J_{N,r}\\
	J_{N,r} &= \sum_{i=r}^{N}(-1)^{i-r}h_{i}(x_i,x_{i-1}+x_i,\dots,x_1+x_2+\cdots+x_i)\nonumber\\
	&\quad\times h_{N-i}(x_{i+1},x_{i+1}+x_{i+2},\dots,x_{i+1}+x_{i+2}+\cdots+x_{N})\\
	D_N &= \sum_{r=1}^{N}(-1)^{r-1}x_r D_{N,r}\\
	D_{N,r} &= \sum_{i=r}^{N}(-1)^{i-r}\left(\frac{1}{x_i}\frac{1}{x_{i-1}+x_i}\cdots\frac{1}{x_1+x_2+\cdots+x_i}\right)\nonumber\\
	&\quad\times\left(\frac{1}{x_{i+1}+x_{i+2}+\cdots+x_{N}}\frac{1}{x_{i+1}+x_{i+2}+\cdots+x_{N-1}}\cdots \frac{1}{x_{i+1}}\right)
\end{align*}
where $D$ is referred to as the denominator analogue of $J$.

The physical logic to the transformation involves obtaining a sequence of denominators which reduce by one in length each time. In the partitioned product take the two middle denominators and merge them using the denominator identity. Extract the longest denominator and what is left is a sum of two pieces with one fewer denominator each than the original. Then repeat the procedure on the new middle pair of denominators which gives an extracted denominator with one fewer element. This procedure provides a linear combination of strings of denominators which reduce in size. Once one of the two partitioned pieces becomes the identity then the other fills out the product so that it ends with a single element. The remaining issues are to process indices using permutations which is the rest of this appendix.

First we recall some permutation notation from the main text and start by redefining the symbol
\begin{align*}
	[x_1x_2\cdots x_n] \equiv x_1\frac{1}{x_1}\frac{1}{x_1+x_2}\cdots\frac{1}{x_1+x_2+\cdots+x_{n}}.
\end{align*}
We also require the previous reversal operators, the first few act on this symbol as
\begin{align*}
	R_2[x_1x_2] &= [x_2x_1]\\
	R_3[x_1x_2x_3] &= [x_3x_2x_1]\\
	R_4[x_1x_2x_3x_4] &= [x_4x_3x_2x_1]\\
	R_4[x_1x_2x_3] &= [x_4x_3x_2]
\end{align*}
where the final line shows how to deal with $R_{N}[x_1\cdots x_{N-1}]$ which permutes the variables in the same way, leaving out the final one; this is also relevant in the main text.

We are going to prove that the denominator analogue resums into a linear combination of permutation operators applied to a string of reciprocals,
\begin{align*}
	x_rD_{N,r}\mapsto S_{N,r}[x_1x_2\cdots x_{N}]
\end{align*}
where the permutations are provided by
\begin{align*}
	S_{N,r} &= S_{N-1,r}+R_{N}R_{N-1}S_{N-1,r-1},\quad 1\leq r\leq N,\quad N>1\\
	S_{N,0}&=0=S_{N,N+1},\quad N>1,\quad S_{1,1}\equiv 1.
\end{align*}
This is the same permutation structure as the coefficient of  $(-1)^{r-1}x_r$ in the hyperbolic permutation representation.

We will use the inductive assumption that 
\begin{align*}
	x_r D_{N-1,r}=S_{N-1,r}[x_1x_2\cdots x_{N-1}]
\end{align*}
and first prove the base case along with a non-trivial example to show the recursive cancellation taking place.
The first example is $x_1D_{1,1}=x_1\frac{1}{x_1}=S_{1,1}[x_1]$ which is trivial. Next $x_2D_{2,2}=[x_2x_1]=R_2[x_1x_2]=S_{2,2}[x_1x_2]$ which is again trivial. Next
\begin{align*}
	D_{2,1}&=\frac{1}{x_1}\frac{1}{x_2}-\frac{1}{x_2}\frac{1}{x_1+x_2}=\left(\frac{1}{x_1}+\frac{1}{x_2}\right)\frac{1}{x_1+x_2}-\frac{1}{x_2}\frac{1}{x_1+x_2}=\frac{1}{x_1}\frac{1}{x_1+x_2}\\
	x_1D_{2,1}&=S_{2,1}[x_1x_2]
\end{align*}
where we used the fundamental denominator identity on the first term on the right-hand side of the first line.

The first example that provides non-trivial permutations occurs in
\begin{align*}
	D_{3,2} &= \frac{1}{x_2}\frac{1}{x_1+x_2}\frac{1}{x_3}-\frac{1}{x_3}\frac{1}{x_2+x_3}\frac{1}{x_1+x_2+x_3}\nonumber\\
	&=\frac{1}{x_2}\left[\frac{1}{x_1+x_2}+\frac{1}{x_3}\right]\frac{1}{x_1+x_2+x_3}-\frac{1}{x_3}\frac{1}{x_2+x_3}\frac{1}{x_1+x_2+x_3}\nonumber\\
	&=\left[\frac{1}{x_2}\frac{1}{x_1+x_2}+\frac{1}{x_2}\frac{1}{x_3}-\frac{1}{x_3}\frac{1}{x_2+x_3}\right]\frac{1}{x_1+x_2+x_3}=\left[D_{2,2}+R_3R_2 D_{2,1}\right]\frac{1}{x_1+x_2+x_3}
\end{align*}
where we used the fundamental denominator identity on the first term when going from the first to the second line, $R_3R_2$ provides a cyclic permutation and we have already solved the previous cases to provide
\begin{align*}
	x_2D_{3,2} &=\left(S_{2,2}+R_2R_1 S_{2,1}\right)[x_1x_2]\frac{1}{x_1+x_2+x_3}=\left(R_2+R_3R_2 \right)[x_1x_2x_3]=S_{3,2}[x_1x_2x_3].
\end{align*}

Now we prove the general case. Consider $x_rD_{M,r}$, we first use the fundamental denominator identity at the partition
\begin{align*}
	\frac{1}{x_1+\cdots+x_i}\frac{1}{x_{i+1}+\cdots+x_{M}}=\left(\frac{1}{x_1+\cdots+x_i}+\frac{1}{x_{i+1}+\cdots+x_{M}}\right)\frac{1}{x_1+\cdots+x_{M}}
\end{align*}
and extract the final reciprocal to the end of the product. Of the remaining terms, the first bracketed term simply loses $x_{M}$ to the final reciprocal and has the structure $a_ib_{M-i}\mapsto a_ib_{M-1-i}$; the second term loses $x_1$ into the final reciprocal and has the structure $a_ib_{M-i}\mapsto a_{i-1}b_{M-i}$. However in the second term we need to relabel the parameters by $x_2\mapsto x_1$, $x_3\mapsto x_2$ and so on until $x_1\mapsto x_M$ and is the one lost. This is a cyclic permutation and is generated by the reversals $R_MR_{M-1}$. 
The permutation aspect stems from the idea that the identity extracts out $a+b$ as a denominator and leaves either $a$ or $b$ behind in the two residual pieces. One of these has the only appearance of the largest values of $r$ amongst the $x_r$ and the other has the only appearance of the lowest value of $r$ with $r=1$ in that piece. The permutation is designed to keep the lowest value of $x_r$ for both by relabelling the indices for the other term. This is equivalent to the use of $\sop^p\cdot\sop^{-p}$ in the main text to accomplish the same mathematical task.
The permutation structure is mathematically controlled by 
\begin{align*}
	S_{M,i} &= S_{M-1,i}+R_MR_{M-1}S_{M-1,i-1}
\end{align*}
and we solve this relationship using the generating function
\begin{align*}
	P_N(z) &= -\sum_{r=0}^{N}z^{r}S_{N,r}\\
	P_N(z) &= (1+z R_NR_{N-1})P_{N-1}(z)=\prod_{r=1}^{N-1}(1+z R_{r+1}R_r)
\end{align*}
since $P_1\equiv S_{1,1}\equiv 1$, the product is taken in order of increasing $r$ from right to left, for example $P_3(z)=(1+zR_3R_2)(1+zR_2R_1)$. We want $z=-1$ for which the identity 
\begin{align*}
	(1+xR_{r+1}R_r)(1+yR_r)=(1+xyR_{r+1})(1+yR_r)
\end{align*}
for $y^2=-1$ allows the recursive proof that
\begin{align*}
	P_N(-1) &= \prod_{r=1}^{N-1}(1+(-1)^{r+1}R_{r+1})
\end{align*}
and is seeded by the fact that $R_1\equiv 1$, which we encountered in equation~\eqref{eq:pn-intro} in the main text.

For a generic term, $D_{N,r}$, using the fundamental identity once provides two sets of terms, $D_{N-1,r}$ and $R_{N-1}R_{N-2}D_{N-1,r-1}$ respectively, multiplied by the final reciprocal,
\begin{align*}
	D_{N,r} &= \left(D_{N-1,r}+R_{N-1}R_{N-2}D_{N-1,r-1}\right)\frac{1}{x_1+x_2+\cdots+x_{N}}.
\end{align*}
Now using the inductive assumption we find
\begin{align*}
	x_rD_{N,r} &= \left(S_{N-1,r}+R_{N-1}R_{N-2}S_{N-1,r-1}\right)[x_1x_2\cdots x_{N-1}]\frac{1}{x_1+x_2+\cdots+x_{N}}\\
	&=S_{N,r}[x_1x_2\cdots x_{N}]
\end{align*}
which completes the proof. The total result of this is $D_N=P_N[x_1x_2\cdots x_{N}]$.

This results in the following for the first few order of the denominator analogue
\begin{align*}
	J_2&\mapsto D_2=[x_1x_2]-[x_2x_1]=P_2[x_1x_2]\\
	J_3&\mapsto D_3=[x_1x_2x_3]-\left([x_2x_1x_3]+[x_2x_3x_1]\right)+[x_3x_2x_1]=P_3[x_1x_2x_3]\\
	J_4&\mapsto D_4= [x_1x_2x_3x_4]-\left([x_2x_3x_4x_1]+[x_2x_3x_1x_4]+[x_2x_1x_3x_4]\right)\nonumber\\
	&\qquad+\left([x_3x_4x_2x_1]+[x_3x_2x_4x_1]+[x_3x_2x_1x_4]\right)-[x_4x_3x_2x_1]=P_4[x_1x_2x_3x_4].
\end{align*}

\section{Overcomplete representation}\label{app:overcomplete}
In this appendix we provide an alternative representation of $G_N$ where the arguments are $L_m= x_m-x_{m-1}$, and there are $N+1$ $x_m$'s which appear to over-specify $G_N$, we comment on this later. 

The physical motivation for this appendix is doing perturbation theory for operators related by the BCH formula. When doing perturbation theory there is an exactly solved, diagonal operator, $A$, and a non-diagonal perturbation, $B$. In this situation the arguments to $G_N$ become differences in the eigenvalues of $A$, for example $\mel*{n}{g(L)B}{n'}=g(a_n-a_{n'})\mel*{n}{B}{n'}$, where $g$ is some Taylor-expandable function, $a_n\delta_{nn'}\equiv A_{nn'}$ are the eigenvalues of $A$ and we used $LB\equiv [A,B]$.

To find the overcomplete representation we use the original one in equation~\eqref{eq:rev-init}, identity~(5.2) from the previous paper, provided in equation~\eqref{eq:ml-5.2}, and its reversed analogue that includes $L_{N+1}$ at the expense of $L_1$,
\begin{align}
	s_N &= f_N(-L_{N+1},-L_N,\dots,-L_2)\nonumber\\
	&\quad+\sum_{r'=1}^{N}f_{r'-1}(-L_{r'},\dots,-L_2)f_1(L_{r'+1}+\cdots+L_{N+1})f_{N-r'}(L_{r'+1},\dots,L_N)\label{eq:ml-5.2-rev},
\end{align}
where $r+r'=N+1$ controls the reversal structure.
We also use $h_r$ in favour of $f_r$ (these quantities are related as in equation~\eqref{eq:fh}) in the final result of this appendix since $h_r$ forms a basis that is analogous to the denominators in ordinary perturbation theory, and that is the physical motivation for this appendix.

First we let
\begin{align*}
	L_1=x_1-x_0,\quad L_2=x_2-x_1,\quad\cdots,\nonumber\\
	L_{N}=x_{N}-x_{N-1},\quad L_{N+1}=x_{N+1}-x_{N}
\end{align*}
which amounts to all commutator operators becoming differences in eigenvalues of the diagonal operator, and then
$$L_1+L_2+\cdots+L_r=x_r-x_0$$
 and 
 $$L_{r+1}+L_{r+2}+\cdots+L_{N+1}=x_{N+1}-x_{r}.$$ 
With this change of variables, equation~\eqref{eq:gn} becomes
\begin{align}
	G_{N+1}(x_1-x_0,x_2-x_1,&\dots,x_{N+1}-x_N)=G_1(x_{N+1}-x_0)\bigg(s_N\nonumber\\
	&+\sum_{r=1}^{N}\frac{(x_0-x_r)h_1(x_0-x_r)-(x_{N+1}-x_r)h_1(x_{N+1}-x_r)}{x_{N+1}-x_0}\nonumber\\
	&\times h_{r-1}(x_{r-1}-x_r,\dots,x_1-x_r)h_{N-r}(x_{r+1}-x_r,\dots,x_{N}-x_{r})\bigg),\label{eq:rev-init}
\end{align}
and the two identities~\eqref{eq:ml-5.2} and~\eqref{eq:ml-5.2-rev} become
\begin{align}
	s_N &=h_N(x_1-x_0,\dots,x_N-x_0)\nonumber\\
	&\quad+\sum_{r=1}^{N}h_{r-1}(x_{r-1}-x_r,\dots,x_1-x_r)h_1(x_0-x_r)h_{N-r}(x_{r+1}-x_r,\dots,x_N-x_r),\label{eq:ml-5.2-h}\\
	s_N&=h_N(x_N-x_{N+1},\dots,x_1-x_{N+1})\nonumber\\
	&\quad+\sum_{r'=1}^{N}h_{r'-1}(x_{r'-1}-x_{r'},\dots,x_1-x_{r'})h_1(x_{N+1}-x_{r'}) h_{N-r'}(x_{r'+1}-x_{r'},\dots,x_N-x_{r'})\label{eq:ml-5.2-rev-h}
\end{align}
respectively, in terms of $h_r$, where we used that $N$ is even to simplify the second identity.
In equation~\eqref{eq:rev-init}, we can recognise the piece with $x_0$ as identity~\eqref{eq:ml-5.2-h} and rewrite as
\begin{align*}
	-\frac{x_0}{x_{N+1}-x_0}h_{N}(x_1-x_0,x_2-x_0,\dots,x_N-x_{0})G_1(x_{N+1}-x_0)
\end{align*}
where we also used the symmetry given in equation~\eqref{eq:hn-sym} to rewrite the arguments of the resulting $h_N$ in ascending order. For the piece with $x_{N+1}$ we can recognise the reversed identity~\eqref{eq:ml-5.2-rev-h} and rewrite as
\begin{align*}
	\frac{x_{N+1}}{x_{N+1}-x_0}G_1(x_{N+1}-x_0)h_{N}(x_1-x_{N+1},x_2-x_{N+1},\dots,x_N-x_{N+1}),
\end{align*}
where we cancelled the $s_N$ terms with those present in equation~\eqref{eq:rev-init}. For the generic piece with $x_r$, $1<r<N+1$ we can use a fundamental hyperbolic fact that appears as identity~(5.1) in the previous paper,
\begin{align*}
	\frac{x_r(-\coth(x_0-x_r)+\coth(x_{N+1}-x_r))}{\sinh(x_{N+1}-x_0)}&=\frac{x_r}{\sinh(x_{N+1}-x_r)\sinh(x_r-x_0)}\\
	&=\frac{x_r G_1(x_{N+1}-x_r)G_1(x_r-x_0)}{(x_{N+1}-x_r)(x_r-x_0)}.
\end{align*}
This way of specifying $G_N$ is subtle, as at first they appear over-specified. The arguments of $G_N$ involve $N+1$ parameters. If one adds a constant, $\lambda$ say, to all the variables then the left-hand side is unchanged but a term proportional to $\lambda$ appears on the right-hand side. This additional term vanishes as explained in this appendix. The structure has been chosen to highlight the structure relevant to physical applications when $A$ is diagonal. The resulting formula is
\begin{align}
	G_{N+1}(x_1-x_0,x_2-x_1,&\dots,x_{N+1}-x_N)=\frac{x_{N+1}}{x_{N+1}-x_0}G_1(x_0-x_{N+1})\nonumber\\
	&\times h_N(x_1-x_{N+1},\dots,x_N-x_{N+1})\nonumber\\
	&+\sum_{r=1}^{N}\bigg[\frac{x_r^2}{(x_r-x_{0})(x_{N+1}-x_r)}G_1(x_0-x_r)h_{r-1}(x_1-x_r,\dots,x_{r-1}-x_r)\frac{1}{x_r}\nonumber\\
	&\qquad\qquad\times h_{N-r}(x_{r+1}-x_r,\dots,x_N-x_r)G_1(x_{N+1}-x_r)\bigg]\nonumber\\
	&\qquad\qquad-\frac{x_0}{x_{N+1}-x_0}h_N(x_1-x_0,\dots,x_N-x_0)G_1(x_{N+1}-x_0),
\end{align}
a few low-order examples are
\begin{equation}\label{eq:g1-to-4-def}
\begin{aligned}
	&G_1(x_1-x_0)={}\frac{x_1-x_0}{\sinh(x_1-x_0)}\\
&G_2(x_1-x_0,x_2-x_1) ={}\frac{x_2}{x_2-x_0}G_1(x_0-x_2)h_1(x_1-x_2)-\frac{x_0}{x_2-x_0}h_1(x_1-x_0)G_1(x_2-x_0)\\
&\quad+\frac{x_1^2}{(x_1-x_0)(x_2-x_1)}G_1(x_0-x_1)\frac{1}{x_1}G_1(x_2-x_1)\\
&G_3(x_1-x_0,x_2-x_1,x_3-x_2) ={} \frac{x_3}{x_3-x_0}G_1(x_0-x_3)h_2(x_1-x_3,x_2-x_3)\\
&\quad-\frac{x_0}{x_3-x_0}h_2(x_1-x_0,x_2-x_0)G_1(x_3-x_0)\\
&\quad+\frac{x_1^2}{(x_1-x_0)(x_3-x_1)}G_1(x_0-x_1)\frac{1}{x_1}h_1(x_2-x_1)G_1(x_3-x_1)\\
&\quad+\frac{x_2^2}{(x_2-x_0)(x_3-x_2)}G_1(x_0-x_2)h_1(x_1-x_2)\frac{1}{x_2}G_1(x_3-x_2)\\
&G_4(x_1-x_0,x_2-x_1,x_3-x_2,x_4-x_3) ={}\frac{x_4}{x_4-x_0}G_1(x_0-x_4)h_3(x_1-x_4,x_2-x_4,x_3-x_4)\\
&\quad-\frac{x_0}{x_4-x_0}h_3(x_1-x_0,x_2-x_0,x_3-x_0)G_1(x_4-x_0) \\
&\quad +\frac{x_1^2}{(x_1-x_0)(x_4-x_1)}G_1(x_0-x_1)\frac{1}{x_1}h_2(x_2-x_1,x_3-x_1)G_1(x_4-x_1)\\
&\quad+\frac{x_2^2}{(x_2-x_0)(x_4-x_2)}G_1(x_0-x_2)h_1(x_1-x_2)\frac{1}{x_2}h_1(x_3-x_2)G_1(x_4-x_2)\\
&\quad+\frac{x_3^2}{(x_3-x_0)(x_4-x_3)}G_1(x_0-x_3)h_2(x_1-x_3,x_2-x_3)\frac{1}{x_3}G_1(x_4-x_3)
\end{aligned}
\end{equation}
and a higher order example is
{\small
\begin{align}
	&G_8(x_1-x_0,x_2-x_1,x_3-x_2,x_4-x_3,x_5-x_4,x_6-x_5,x_7-x_6,x_8-x_7) =\nonumber\\ &\frac{x_8}{x_8-x_0}G_1(x_0-x_8)h_7(x_1-x_8,x_2-x_8,x_3-x_8,x_4-x_8,x_5-x_8,x_6-x_8,x_7-x_8)\nonumber\\
	&-\frac{x_0}{x_8-x_0}h_7(x_1-x_0,x_2-x_0,x_3-x_0,x_4-x_0,x_5-x_0,x_6-x_0,x_7-x_0)G_1(x_8-x_0)\nonumber\\
	&+\frac{x_1^2}{(x_1-x_0)(x_8-x_1)}G_1(x_0-x_1)\frac{1}{x_1}h_6(x_2-x_1,x_3-x_1,x_4-x_1,x_5-x_1,x_6-x_1,x_7-x_1)G_1(x_8-x_1)\nonumber\\
	&+\frac{x_2^2}{(x_2-x_0)(x_8-x_2)}G_1(x_0-x_2)h_1(x_1-x_2)\frac{1}{x_2}h_5(x_3-x_2,x_4-x_2,x_5-x_2,x_6-x_2,x_7-x_2)G_1(x_8-x_2)\nonumber\\
	&+\frac{x_3^2}{(x_3-x_0)(x_8-x_3)}G_1(x_0-x_3)h_2(x_1-x_3,x_2-x_3)\frac{1}{x_3}h_4(x_4-x_3,x_5-x_3,x_6-x_3,x_7-x_3)G_1(x_8-x_3)\nonumber\\
	&+\frac{x_4^2}{(x_4-x_0)(x_8-x_4)}G_1(x_0-x_4)h_3(x_1-x_4,x_2-x_4,x_3-x_4)\frac{1}{x_4}h_3(x_5-x_4,x_6-x_4,x_7-x_4)G_1(x_8-x_4)\nonumber\\
	&+\frac{x_5^2}{(x_5-x_0)(x_8-x_5)}G_1(x_0-x_5)h_4(x_1-x_5,x_2-x_5,x_3-x_5,x_4-x_5)\frac{1}{x_5}h_2(x_6-x_5,x_7-x_5)G_1(x_8-x_5)\nonumber\\
	&+\frac{x_6^2}{(x_6-x_0)(x_8-x_6)}G_1(x_0-x_6)h_5(x_1-x_6,x_2-x_6,x_3-x_6,x_4-x_6,x_5-x_6)\frac{1}{x_6}h_1(x_7-x_6)G_1(x_8-x_6)\nonumber\\
	&+\frac{x_7^2}{(x_7-x_0)(x_8-x_7)}G_1(x_0-x_7)h_6(x_1-x_7,x_2-x_7,x_3-x_7,x_4-x_7,x_5-x_7,x_6-x_7)\frac{1}{x_7}G_1(x_8-x_7)\label{eq:g8}.
\end{align}
}
\bibliographystyle{unsrt}
\bibliography{refs}
\end{document}